\documentclass[useAMS]{mn2e}
\onecolumn
\input{psfig}
\usepackage{times}
\usepackage{bm}
\usepackage{amsmath} 
\usepackage{url}
\usepackage{dblfnote}
\usepackage{aas_macros}
\usepackage{hyperref}
\def\Real{{\rm I\mathchoice{\kern-0.70mm}{\kern-0.70mm}{\kern-0.65mm}%
 {\kern-0.50mm}R}}

\font \bolditalics = cmmib10
\def\bx#1{\leavevmode\thinspace\hbox{vrule\vtop{\vbox{\hrule\kern1pt
 \hbox{\vphantom{\tt/}\thinspace{\bf#1}\thinspace}}
 \kern1pt\hrule}\vrule}\thinspace}

\def \vc #1{{\textfont1=\bolditalics \hbox{$\bf#1$}}}

\newcommand{\bea}{\begin{eqnarray}}
\newcommand{\eea}{\end{eqnarray}}

\def\ps{{\bf s}}
\def\xg{{\bf x}}

\def\eg{{\bf e}}

\def\thetag{{\vc \theta}}

\def\varthetag{{\vc \vartheta}}
\def\eg{{\vc e}}
\def\gammag{{\vc \gamma}}

\def\be{\begin{equation}}
\def\ee{\end{equation}}

\def\ba{\begin{eqnarray}}
\def\ea{\end{eqnarray}}

\def \vc #1{{\textfont1=\bolditalics \hbox{$\bf#1$}}}{\catcode`\@=11

\def\ltsima{$\; \buildrel < \over \sim \;$}
\def\lsim{\lower.5ex\hbox{\ltsima}}
\def\gtsima{$\; \buildrel > \over \sim \;$}
\def\gsim{\lower.5ex\hbox{\gtsima}}

\title[Measurement of three-point shear statistics with COSMOS]{Weak lensing from space\thanks{Based on observations made with the NASA/ESA Hubble Space Telescope, obtained from the data archives at the Space Telescope European Coordinating Facility and the Space Telescope Science Institute, which is operated by the Association of the Universities for Research in Astronomy, Inc., under NASA contract NAS 5-26555.}: first cosmological constraints from three-point shear statistics.}
\author[Semboloni et al.]{Elisabetta Semboloni$^{1,2}$\thanks{sembolon@strw.leidenuniv.nl}, Tim Schrabback$^{1,2}$, Ludovic van Waerbeke$^3$, Sanaz Vafaei$^3$, \newauthor Jan Hartlap$^2$, Stefan Hilbert$^{2,4}$ \\
$^1$Leiden Observatory, Leiden University, NL-2333 CA Leiden, The Netherlands \\
$^2$Argelander-Institut f\"{u}r Astronomie, Auf dem H\"{u}gel 71, D-53121, Bonn, Germany\\
$^3$University of British Columbia, Department of Physics \& Astronomy, 6224
Agricultural Road, Vancouver, B.C. V6T 1Z1, Canada \\
$^4$Max Planck Institut f\"{u}r Astrophysik, Karl-Schwarzschild-Str. 1, 85741 
Garching, Germany\\}
\begin{document}
\maketitle
\begin{abstract}
 We use weak lensing data from the {\it Hubble Space Telescope} COSMOS survey to measure the second- and third-moments of the cosmic shear field, estimated  from about $450\,000$ galaxies with average redshift $\bar z \sim 1.3$. 

We measure two- and three-point shear statistics using a tree-code, dividing the signal in E, B and mixed components. We present a detection of the third-order moment of the aperture mass statistic and verify that the measurement is robust against systematic errors caused by  point spread function (PSF) residuals and by the intrinsic alignments between galaxies. The amplitude of the measured three-point cosmic shear signal is in very good agreement with the predictions for a WMAP7 best-fit model, whereas the amplitudes of potential systematics are consistent with zero.

We make use of three sets of large ${\rm \Lambda CDM}$  simulations to test the accuracy of the cosmological predictions and to estimate the influence of the cosmology-dependent covariance.

We perform a likelihood analysis using the measurement of $\langle M_{\rm ap}^3\rangle(\theta)$ and find that the $\Omega_m -\sigma_8$ degeneracy direction is well fitted by the relation:
 \mbox{$\sigma_8 (\Omega_m/0.30)^{0.49}=0.78^{+0.11}_{-0.26}$} which is in good agreement with the best fit relation obtained by using the measurement of $\langle M_{\rm ap}^2\rangle(\theta)$: \mbox{$\sigma_8 (\Omega_m /0.30 )^{0.67}=0.70^{+0.11}_{-0.14}$}.

 We present the first measurement of the more generalised three-point shear statistic $\langle M_{\rm ap}^3\rangle(\theta_1,\theta_2,\theta_3)$ and  find a very good agreement with the WMAP7 best-fit cosmology.  The cosmological interpretation of $\langle M_{\rm ap}^3\rangle(\theta_1,\theta_2,\theta_3)$ gives \mbox{$\sigma_8 (\Omega_m/0.30)^{0.46}=0.69 ^ {+0.08}_{-0.14}$}. Furthermore,  the combined likelihood analysis of $\langle M_{\rm ap}^3\rangle(\theta_1,\theta_2,\theta_3)$ and $\langle M_{\rm ap}^2\rangle(\theta)$ improves the accuracy of the cosmological constraints to \mbox{$\sigma_8 (\Omega_m/0.30)^{0.50}=0.69 ^ {+0.07}_{-0.12}$}, showing the high potential of this combination of measurements to infer cosmological constraints.
\end{abstract}


\section{Introduction}
Weak lensing by large-scale structure, i.e. cosmic shear, constitutes a powerful tool to investigate the statistical properties of the dark matter
distribution. In the past years, the measurement of the observed correlation
between pairs of galaxies has been successfully used to constrain the power
spectrum of matter fluctuations (see Benjamin et al. 2007, Fu et al. 2008 and Schrabback et al. 2010 for the most recent results). With the amount of data available from current and future surveys, we should in principle be able to measure higher order
statistics which provide a measurement of the non-Gaussian features  of the matter distribution. In particular, third-order statistics are expected to
increase the strength of the constraints obtained by weak lensing studies \cite{BvWM97,vWetal99,vWetal01,KiSc05,TaJa05,Vaetal09,Beetal10}.

However, to date, predictions for three-point shear statistics are still
inaccurate, as they are based on perturbative development and rely on the
non-linear evolution of the power spectrum which is generally approximated by empirical models. 
Moreover, one needs to assess the accuracy of the measurement of the cosmic shear statistics in order to infer unbiased constraints on the cosmological parameters. 
In the past years, various publications focused on the study of two types of systematic effects which are likely to affect cosmic shear measurements. The first one is a measurement bias due to imperfections in the methods used to measure the ellipticity of galaxies and correct for the image point-spread-function (PSF) \cite{Heetal06a,Maetal07a,Bretal09,Kietal08,Seetal09}. The second one arises from the fact that galaxies align according to the local gravitational field \cite{Cretal01,HiSe04} creating intrinsic correlations.

 All these complications might explain why so far there have been only three attempts to
 measure three-point shear statistics: two using VIRMOS data 
 \cite{Beetal02,Peetal03} and one using CTIO data \cite{Jaetal04}.
Due to issues mentioned above, neither of these papers presented a cosmological interpretation.

Since then, there has been significant progress in the field and today we are able to quantify the impact of bias on the measurement of cosmic shear statistics, using simulations and various new diagnostics. Furthermore, the quality of the data has improved as well as our ability to interpret the measurement.

In this paper, we present a measurement of three-point shear statistics obtained using the {\it Hubble Space Telescope} COSMOS data. We study the reliability of this measurement against systematics and use it to infer constraints on cosmological parameters.

The COSMOS survey \cite{Scoetal07a} combines high-resolution HST/ACS imaging covering a total area of 1.64 ${\rm deg}^2$ \cite{Scoetal07b} with deep ground-based  multi-colour data, providing accurate photometric redshifts \cite{Iletal09}. Hence, this field provides a unique set of data to study the formation of structures and is particularly well suited for 3D weak lensing studies, such as the measurement of two-point shear statistics in tomographic bins \cite{Maetal07b,Scetal10}.

The paper is organised as follows.  In Section \ref{sec:theory} we introduce the notation and  we test the accuracy of the cosmological predictions using various sets of $N$-body simulations.
In Section \ref{sec:data} we describe the data we use. In Section \ref{sec:measure} we present the measurement of the two- and three-point shear statistics from COSMOS and test its reliability against systematic errors by PSF residuals and intrinsic ellipticity correlations. In Section \ref{sec:covariance} we explain the procedure used to determine the covariance matrix which we need to perform a likelihood analysis of our measurement.
 In Section \ref{sec:constraints}, we perform a likelihood analysis and present the cosmological constraints we obtain.
 Finally, in Section \ref{sec:conclusions} we discuss our results.
\section{Theory}\label{sec:theory}
\subsection{Aperture mass statistics}
In this paper we present measurements of second- and third-order shear
statistics obtained by using an aperture mass filter. Using the same notation as Schneider et al. (1998), the aperture mass is defined by:
\be\label{eq:map_def}
M_{\rm ap} (\theta)=\int d^2 \vartheta U_\theta(\vartheta) \kappa (\varthetag) \equiv \int d^2 \vartheta Q_\theta(\vartheta) \gamma_t (\varthetag)\,, 
\ee
where $U_\theta$ is an axially symmetric filter of characteristic size $\theta$
with zero mean. The second equality can be derived using the relation
between the projected density field $\kappa$ and the shear field $\vc
\gamma=\gamma_t+i\gamma_\times$. We define the tangential $\gamma_t$ and the cross $\gamma_\times$ shear components for a galaxy at the position $\varthetag$, as the projections of $\vc \gamma$ parallel and $45^\circ$ rotated with respect to the line connecting the centre of the aperture and the centre of the galaxy.

 Aperture mass statistics are sensitive only to the E-modes of the observed shear field \cite{Cretal02}. Replacing the tangential shear $\gamma_t$ in Equation~(\ref{eq:map_def}) with the cross component $\gamma_\times$ one measures the non-gravitational, B-mode component of the observed shear field, $M_{\rm \times} (\theta)$. 
 A filter which is able to separate E- and B-modes unambiguously has a noticeable advantage when compared to other types of filters as this division can be used to test for the presence of systematics affecting the measurement of cosmic shear statistics.
 
The relation between the two-point shear statistic $\langle M_{\rm
 ap}^2\rangle(\theta)$ for a given characteristic scale $\theta$
and the underlying cosmology is expressed by \cite{Ka98,Scetal98}:
\be\label{eq:map2}
 \langle M_{\rm ap}^2\rangle(\theta)={2\pi}\frac{9H_0^4}{4c^4} \Omega_{ m}^2 \int^{w_H}_{0} dw \frac{g^2(w)}{a^2(w)}
 \int_0^\infty s ds P\Big(\frac{s}{f_\mathrm{k}(w)},w\Big)[I(s\theta)]^2\;,
\ee
with 
\be\label{eq:distr}
g(w)=\int_w^{w_H} dw^\prime p_s(w^\prime) \frac{f_\mathrm{k}(w-w^\prime)}{f_\mathrm{k}(w^\prime)}\;,
\ee
where $f_\mathrm{k}(w)$ is the comoving angular distance, $w$ is the radial
coordinate, $w_H$ is the radial coordinate of the horizon; $H_0$, $\Omega_{\rm m}$ and $\sigma_8$ are the Hubble constant,
the matter density and the normalisation of the power spectrum of the matter
fluctuations. $\langle M_{\rm ap}^2\rangle(\theta)$ depends on the power spectrum of matter fluctuations \mbox{$P\Big(\frac{s}{f_\mathrm{k}(w)},w\Big)$} and on the distribution of the sources $ p_{\rm s}(w)$.
Finally $I(s \theta)$ is the Fourier transform of the filter $U_\theta(\vartheta)$.

A similar relation can be derived for $\langle M_{\rm ap}^3\rangle(\theta)$ in the perturbative regime \cite{Scetal98}:
\ba\label{eq:map3}
 \langle M_{\rm ap}^3\rangle(\theta)=\frac{81H_0^6 \Omega_{m}^3}{8\pi c^6}\int^{w_H}_{0} dw \frac{g^3(w)}{a^3(w)f_\mathrm{k}(w)}
\int_0^\infty d^2 \ps_1 P\Big(\frac{s_1}{f_\mathrm{k}(w)},w\Big)I(s_1\theta)
 \int_0^\infty d^2 \ps_2 P\Big(\frac{s_2}{f_\mathrm{k}(w)},w\Big)I(s_2\theta)
 I(|\ps_1+\ps_2|\theta) F_2(\ps_1,\ps_2)\;
\ea
 where $F_2(\ps_1,\ps_2)$ represents the
coupling between two density fluctuation modes characterised by
the wave vectors $\ps_1$ and $\ps_2$. 

One can estimate the value of $M_{\rm ap}({\theta})$ from the shear field as stated by Equation (\ref{eq:map_def}), replacing the integral by a finite sum over the galaxies in the field. One can then use this measurement to estimate the second- and third-order moments of the aperture mass. However, on real data, this direct measurement does not allow one to separate the E- and B-modes unambiguously, as the properties of the filter $Q_\theta(\vartheta)$ are in principle lost due to the presence of masked regions. One can overcome this problem by estimating $\langle M_{\rm ap}^2\rangle(\theta)$ and $\langle M_{\rm ap}^3\rangle(\theta)$ using two- and three-point correlation functions as we will explain in more detail in Section \ref{sub:measure_corr}.

 %Moreover, Equation \ref{eq:filter} defines a very narrow window filter, which for a given $\vartheta$ probes essentially modes with $s \sim 2/ \vartheta $.
 \begin{figure*}
\begin{tabular}{ll}
\psfig{figure=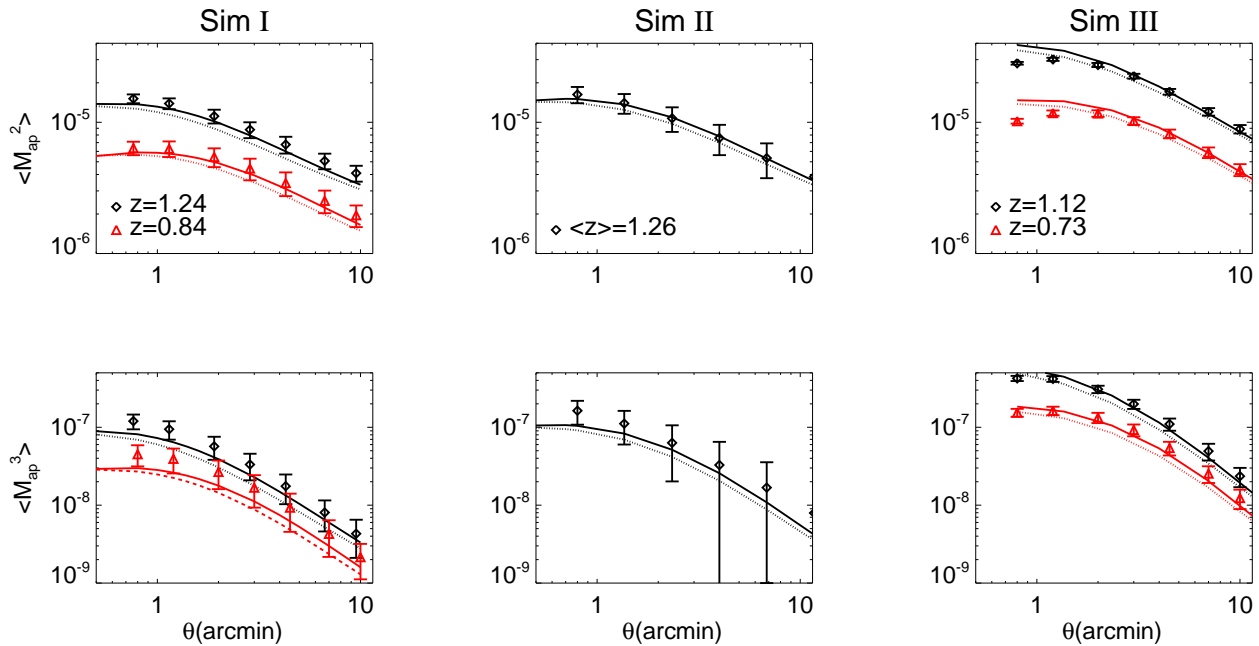,width=.99\textwidth}
\end{tabular}
\caption{\label{fig:test_model} The top plots show the measurement of $\langle M_{\rm ap}^2\rangle(\theta)$ for  the set of simulations I (left panel): $\Omega_m=0.24$, $\sigma_8=0.74$ , II (middle panel): $\Omega_m=0.25$, $\sigma_8=0.9$ and III (right panel): $\Omega_m=0.3$, $\sigma_8=1$.
 For set I we show the signal for two source redshift planes: $z=1.24$ (black diamonds) and $z=0.84$ (red triangles);  for set II we show the signal for a source redshift distribution with $\langle z\rangle=1.26$ (black diamonds). For the the third set of simulations we show  the signal for two source redshift planes: $z=1.12$ (black diamonds) and  $z=0.73$ (red triangles).
The three bottom plots show the measurement of $\langle M_{\rm ap}^3\rangle(\theta)$ for the sets I (left panel), II (middle panel) and III (right panel) and for the same source redshifts.
 The theoretical predictions have been obtained using Equation (\ref{eq:map2}) for $\langle M_{\rm ap}^2\rangle(\theta)$ and Equation (\ref{eq:map3}) for $\langle M_{\rm ap}^3\rangle(\theta)$ for the same cosmology used to produce the simulations. The solid lines show the results obtained using the halofit model to compute the non-linear evolution of the power spectrum and the bispectrum. The dashed lines show the results using the non-linear power spectrum by  PD96.
The error-bars represent the dispersion between all the lines-of-sight available for each set of simulations. }
\end{figure*}

\subsection{Testing the accuracy of the cosmological predictions}\label{sec:model}
In this section we investigate the accuracy with which $\langle M_{\rm ap}^3\rangle(\theta)$ can be predicted using Equation (\ref{eq:map3}). Indeed, as we already pointed out, this equation is valid only in the perturbative regime so its
precision is expected to decrease at small scales. We compute $F_2(\ps_1,\ps_2)$ using the fitting formula suggested by Scoccimarro \& Couchman (2001).  Scoccimarro \& Couchman (2001)  used  N-body simulations  from various ${\rm \Lambda CDM}$ cosmologies, and found  that this fitting formula is able to predict the bispectrum within $15\%$ accuracy. Van Waerbeke et al. (2001) used this same fitting formula to build third-order weak lensing predictions and found a discrepancy with simulations between $10\%$ and $20\%$.

The accuracy of the fitting formula is scale-dependent, thus we expect weak lensing predictions to have an accuracy which changes as a function of redshift, scale and filter type. For this reason, we directly compare   $\langle M_{\rm ap}^3\rangle(\theta)$ predictions  with values  measured on three very large sets of ray-tracing simulations. All of them use a ${\rm \Lambda CDM}$ cosmology, and the main parameters of each set are summarised below:
\begin{enumerate}

\item The first set of simulations has been described by Vafaei et al. (2010); it is composed of 60 quasi-independent lines-of-sight, each having an area of $12.84~{\rm deg^2}$. The cosmological model is very similar to the best-fit of the WMAP3 data \cite{Spetal07}: $\Omega_\Lambda=0.76$, $\Omega_m=0.24$, $\Omega_b=0.04$, $h=0.7$ and \mbox{$\sigma_8=0.74$}. For each line-of-sight we have 40 convergence maps $\kappa(z_i)$ corresponding each to a source redshift plane $z_i$. Each map contains $1024\times1024$ pixels of angular size $0.21$ arcmin and all together the 40 maps cover a source redshift range between zero and three. 
\item As  second set, we use the Millennium Simulation \cite{Spetal05,Hietal09}. The cosmology is characterised by $\Omega_\Lambda=0.75$, $\Omega_m=0.25$, $\Omega_b=0.045$, $h=0.73$ and \mbox{$\sigma_8=0.9$}. The set  consists of  32 lines-of-sight of $16\, {\rm deg^2}$ each. Each map is $4096\times4096$ pixels of angular size $0.058$ arcmin. The redshift range is between zero and three. 
\item The third and last set of simulations, is described in Semboloni et al. (2007). The cosmological model is characterised by $\Omega_\Lambda=0.7$, $\Omega_m=0.3$, $h=0.7$ and \mbox{$\sigma_8=1.0$}. The set  includes  64 lines-of-sight of $49\, {\rm deg^2}$ each. Each map is $1024\times1024$ pixels of angular size $0.41$ arcmin. The redshift range is between zero and three. 
\end{enumerate}
In general, we confirm that the fitting formula can reproduce the amplitude of $\langle M_{\rm ap}^3\rangle(\theta)$ measured on simulations with the same degree of accuracy found by van Waerbeke et al. (2001). The goodness of the predictions depends on the angular scale (although for small scales we do expect the measurement to be affected by the finite size of the pixels) and slightly changes with the redshift. In the best cases the difference is about a few percent for $\langle M_{\rm ap}^2\rangle(\theta)$ and ten percent for $\langle M_{\rm ap}^3\rangle(\theta)$. 

We show for example, in the  top panels of Figure \ref{fig:test_model} 
the measurement of $\langle M_{\rm ap}^2\rangle(\theta)$ using  sets I (left panel), II (middle panel)  and III (right panel) for various source redshifts. The error-bars represent the variance over the various lines-of-sight. The results have been compared with the predictions obtained using Equation (\ref{eq:map2}) together with a non-linear power spectrum  computed using the  halofit prescription developed by Smith et al. (2003). Furthermore, we show models obtained  using  the fitting formula by Peacock \& Dodds (1996) (hereafter PD96).
 The bottom panels show the measurement of $\langle M_{\rm ap}^3\rangle(\theta)$ using  sets I (left panel), II (middle panel)  and III (right panel) for the same source redshifts.  The results  are to be compared with the predictions made using Equation (\ref{eq:map3}) using either  halofit (solid lines)  or  PD96 (dashed lines) to compute the non-linear power spectrum.

In all cases $\langle M_{\rm ap}^2\rangle(\theta)$ predictions agrees fairly well with  the measurements,  although the models
suggest  values which are smaller than the measurements, especially for  small angular scales. This is not the case for the set  III and it can be explained by noticing that the pixel size for this set  is too large to allow one to measure correlations at scales smaller than few arcminutes. 

 For the two-point shear statistics the lack of agreement is due to the limited accuracy of the power spectrum in the non-linear regime. In this respect, our results are similar to the ones by Hilbert et al. (2009) who compared the projected power spectrum of matter fluctuations measured on the Millennium simulation with cosmological predictions by PD96 and by Smith et al. (2003), and found that both prescriptions strongly under-predict the amplitude of the power spectrum for small scales.

We find that for $\langle M_{\rm ap}^3\rangle(\theta)$ the disagreement is more significant especially at small scales.
This is not surprising as the predictions depend on the non-linear power spectrum squared and they also rely on the perturbative approximation.  The results obtained with the two methods  are similar although PD96  underestimates the amplitude of the signal more than the halofit model. For this reason, we will use the halofit model throughout  this paper.

Overall, these results confirm an already known problem: in the near future we need to increase the accuracy of the cosmological predictions if we want to increase the accuracy of the cosmological interpretation of future data sets. This is not a problem for the aim of this paper which is meant to be a proof of concept and is still limited by statistical accuracy due to the small area of COSMOS.

\section{The dataset}\label{sec:data}
\begin{figure*}
\begin{tabular}{cc}
\psfig{figure=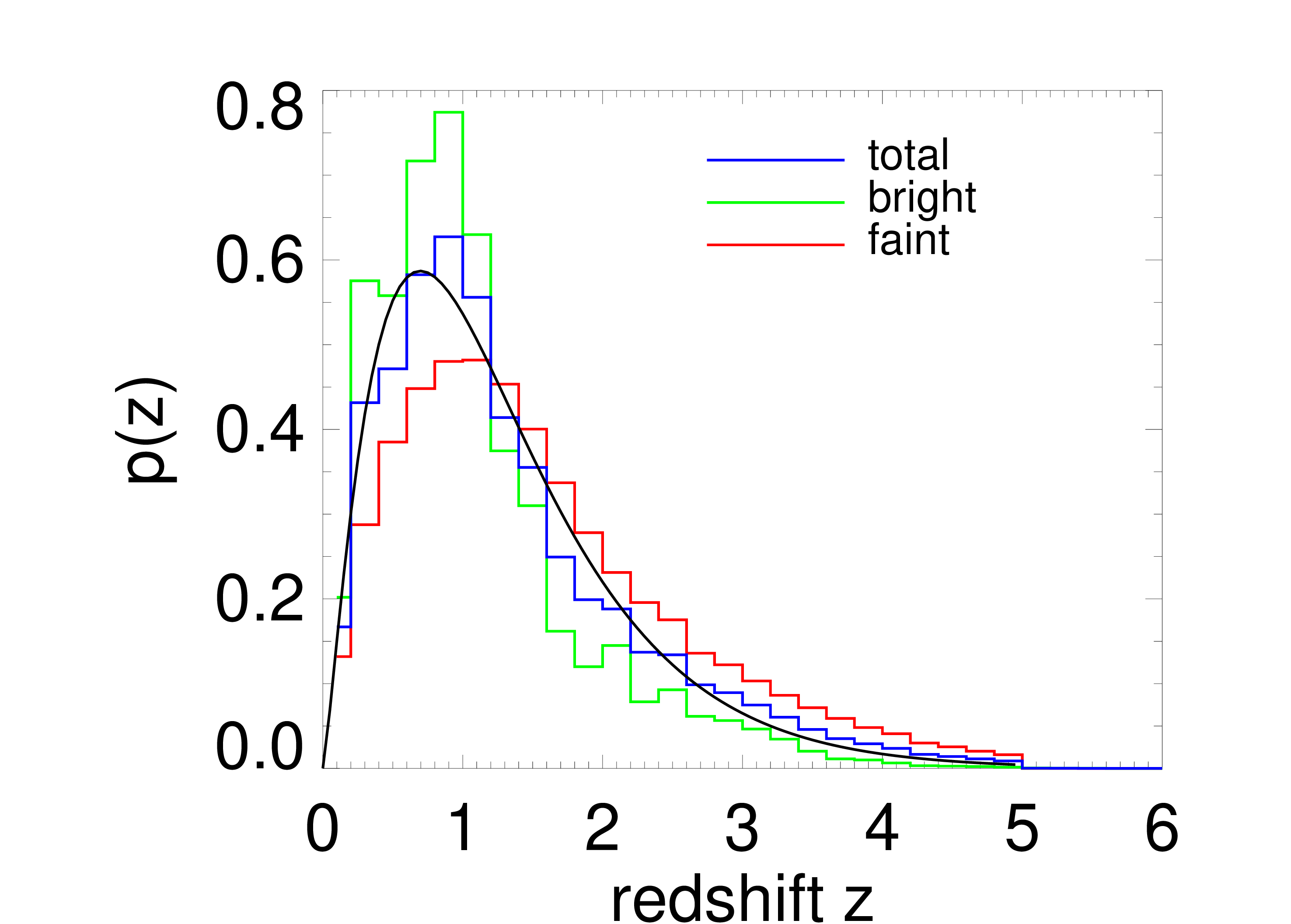,width=.5\textwidth}&\psfig{figure=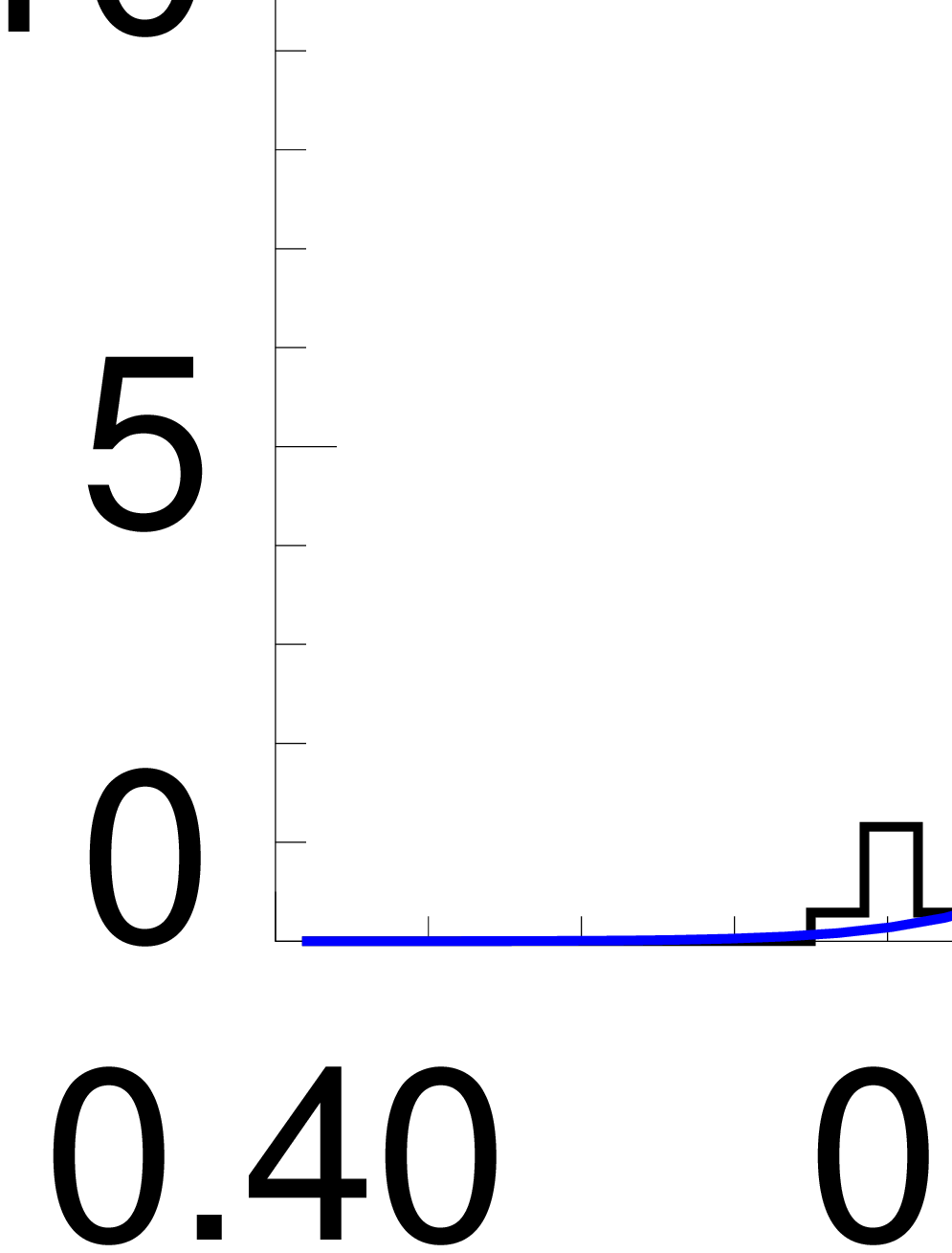,width=.5\textwidth}
\end{tabular}
\caption{\label{fig:histoz}Left panel: the red histogram shows the  estimated redshift distribution of the faint sample.  The green  histogram shows the distribution of the individual photometric redshift estimates in the bright sample. The  blue  histogram shows the redshift distribution of the whole catalogue used to measure the two- and three-point shear statistics. The black line represents the best-fit to the total redshift distribution, obtained by using Equation (\ref{eq:pofz}) which is obtained for $z_0=0.509$, $\alpha=1.29$ and $\beta=0.95$. Right panel: the black histogram shows the probability distribution of the best-fit value  for $z_0$ in Equation (\ref{eq:pofz}) with $\alpha=1.29$ and $\beta=0.95$. The probability distribution has been obtained producing one thousand bootstrap replications accounting for the estimated redshift error (see text). The blue  line shows the Gaussian distribution best-fit of the histogram.}
\end{figure*}
%\begin{figure}
%\psfig{figure=histoz.cosmos.bis.ps,width=.5\textwidth}
%\caption{\label{fig:z0} The black solid histogram shows the probability distribution of the best-fit value  for $z_0$ in Equation (\ref{eq:pofz}) obtained producing one thousand bootstrap replications (see text). The blue solid line shows the Gaussian distribution best-fit of the histogram.}
%\end{figure}

The HST/COSMOS Survey covers 1.64 deg$^2$ consisting of 579 ACS tiles imaged in F814W with an exposure time of
2028s \cite{Scoetal07b}. A detailed description of our COSMOS shear catalogue, together with technical details of its derivation are
given in Schrabback et al. (2010). Here we only briefly summarise its main features.

Our galaxy shape measurements are based on the Erben et al. (2001) implementation of the KSB+ formalism \cite{Kaetal95,LuKa97,Hoetal98} as adapted for space-based
weak lensing measurements in Schrabback et al. (2007). We employ a principal component interpolation for the temporally and 
spatially varying ACS point-spread-function, parametric corrections for charge-transfer-inefficiency for both stars and 
galaxies, and a signal-to-noise dependent shear calibration correction.
We match our shear catalogue to the ground-based photometric redshift catalogue from Ilbert et al. (2009), providing
individual redshift estimates $z_{\rm peak}$ and errors for 194\,976 galaxies with \mbox{$i^+<25$} (Subaru 3'' Sextractor auto-magnitude). We will refer to this subsample as ``bright sample''.
In addition, Schrabback et al. (2010) carefully estimated the redshift distribution $p_{\rm faint}(z)$ for 251\,958 galaxies  with \mbox{$i_{814}<26.7
$} without individual redshifts, with a conservatively estimated 
\mbox{$\sim 10\%$} uncertainty in the redshift calibration. We will refer to this subsample as ``faint sample''. In the left panel of Figure \ref{fig:histoz} we show the normalised redshift histograms for the faint and bright sample. 

In Section \ref{sec:constraints} we will present the likelihood analysis of the second- and third-order moments of the aperture mass measured from the total (faint+bright) sample. The amplitude of the signal depends on the redshift distribution of the sources $p_{\rm s}(z)$, thus the error on $p_{\rm s}(z)$ affects the accuracy of the constraints. We show below  that the error on the redshift distribution can be expressed by a single parameter; this parameter will be varied together with the cosmological parameters and it will act as nuisance parameter.

We derive the redshift distribution of the total sample by adding the histogram of the bright sample and the probability distribution $p_{\rm faint}(z)$ of the faint sample weighted by the number of galaxies contained in each catalogue. We fit the total redshift histogram using the following function \cite{Bretal96}:
\be\label{eq:pofz}
p_{\rm s}(z)=\frac{\beta}{z_0\Gamma \big(\frac{1+\alpha}{\beta}\big)} \Big (\frac{z}{z_0}\Big)^\alpha \exp \Big[ -\Big (\frac{z}{z_0}\Big)^\beta\Big]\,.
\ee
% Since in the fit the dominant source of noise comes from the parametrisation, we do not use a maximum likelihood estimation, which should be more accurate than fitting the histogram but much more time consuming and would not improve the quality of the fit.
In the left panel of Figure \ref{fig:histoz} we show the histogram of the total sample and the best-fit model, which is characterised by $\alpha=1.29$, $\beta=0.95$, $z_0=0.509$ and  average redshift $\bar z=1.27$. 

 We now want to find the error on the parameters $\alpha$, $\beta$ and $z_0$ resulting from the individual redshift errors in the bright sample  and uncertainty in the redshift extrapolation for the faint sample. We create one thousand bootstrap replications of the redshift distribution for the faint sample introducing a nuisance parameter $f_z$ similar to the one used by Schrabback et al. (2010). Using  this parameter we create new redshift distributions $p_{\rm faint}(f_z z)$, selecting $f_z$ from a Gaussian distribution centered at $f_z=1$ and having a dispersion $\sigma_{f_z}=0.10$. 
We create one thousand bootstrap replications for the bright sample selecting the redshift value of each galaxy from a Gaussian distribution with dispersion \mbox{$\sigma_z=(z_{68\%,max}-z_{68\%,min})/2$} where the quantities $z_{68\%,min}$ and $z_{68\%,max}$, define the redshift interval for which the probability distribution is higher than $68\%$. 
We build one thousand replications of the total catalogue by adding a bright sample histogram and a $p_{\rm faint}(z)$ replication of the faint sample.  We notice that $\alpha$ and $\beta$ are  partially  degenerate parameters of the Equation (\ref{eq:pofz}): fixing $\alpha=1.29$ and $\beta=0.95$ and then varying  only $z_0$  allows for a degree of freedom which is high enough to obtain good fits for the various replications. Thus, we determine the best-fit parameter $z_0$ for each replication and build the probability distribution $p(z_0)$. We find that $p(z_0)$ follows fairly well a Gaussian distribution (see right panel of Figure \ref{fig:histoz}) characterised by a standard deviation $\sigma_{z_0}=0.022$, that is about $5\%$ error on the mean redshift. 
 \begin{figure*}
\begin{tabular}{cc}
\psfig{figure=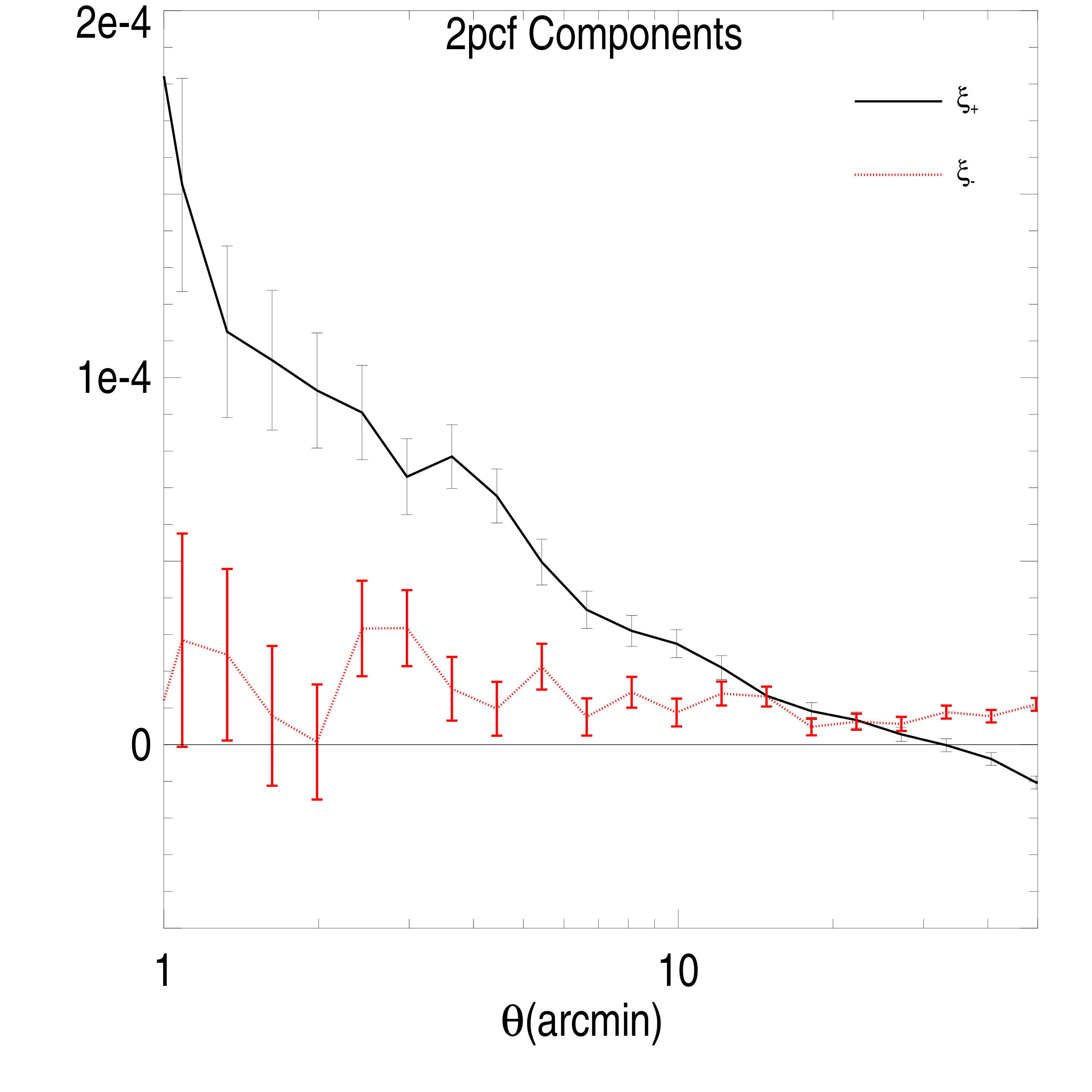,width=.45\textwidth}&\psfig{figure=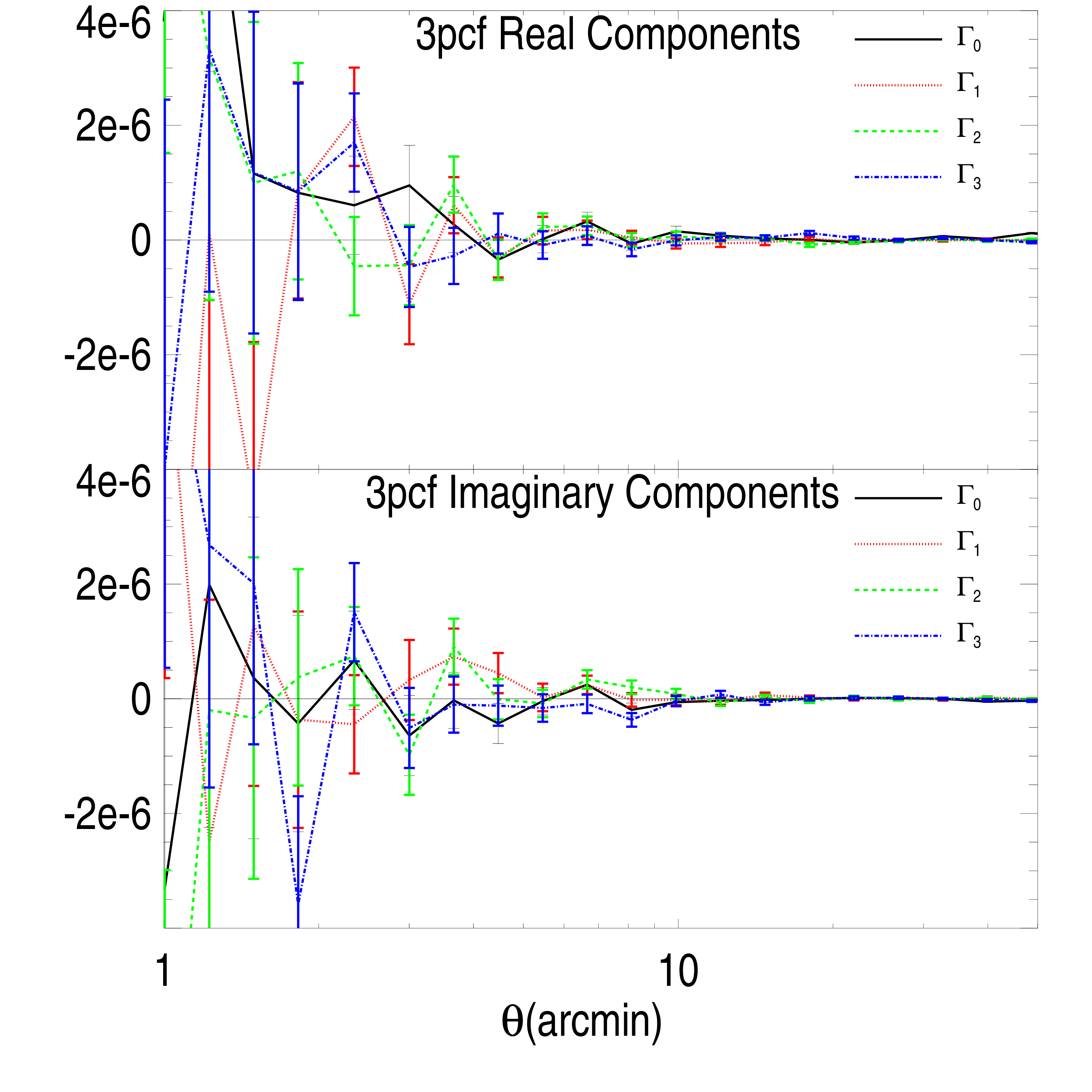,width=.45\textwidth}
\end{tabular}
\caption{\label{fig:correlations} Left panel: two-point correlations $\xi_+(\theta)$ (black solid line)  and  $\xi_-(\theta)$ (red dashed line) measured on the COSMOS galaxy catalogue.  For $\xi_-(\theta)$ we show only the real component as the imaginary component is expected to vanish ($\xi_+(\theta)$ is real by construction). 
Right panel: amplitude of the  three-point correlation functions $\Gamma_i$, measured on quasi-equilateral triangles of side length $\theta$. Because the bins are spaced in a logarithmic way  triangles belonging to the bin  $s=t=|{\bf t} -{\bf s}|=\theta$ are not strictly equilateral.
The top-plot shows the amplitude of the real part, whereas the bottom plot shows the amplitude of the imaginary part. The error-bars are computed as the dispersion measured in each bin and do not include sampling variance.}
\end{figure*}

\section{Measurement and Analysis of systematics}\label{sec:measure}
\subsection{Measurement of two- and three-point shear statistics}\label{sub:measure_corr}
 \begin{figure*}
\begin{tabular}{cc}
\psfig{figure=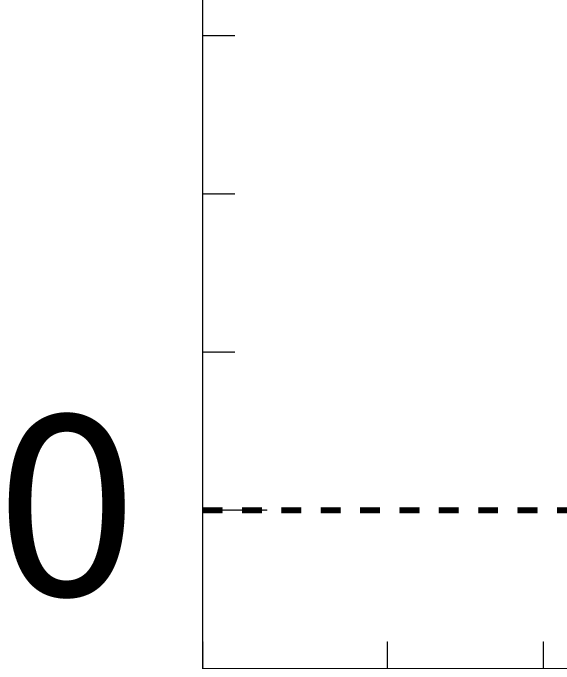,width=.45\textwidth}&\psfig{figure=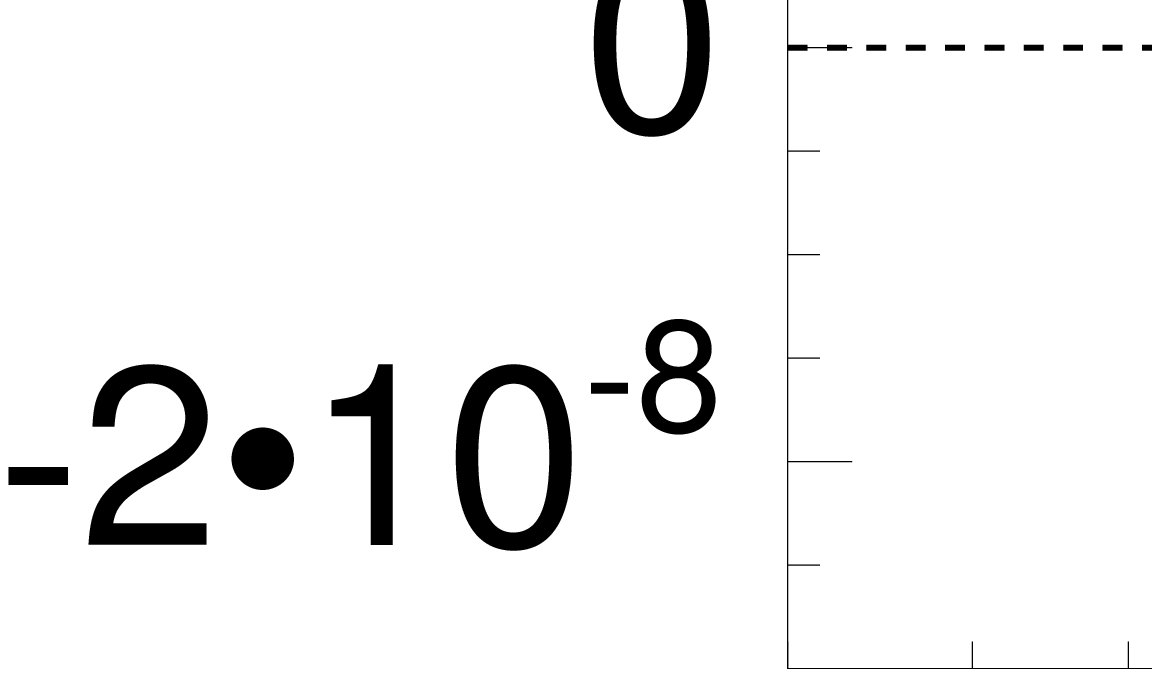,width=.45\textwidth}
\end{tabular}
\caption{\label{fig:map} The left panel shows the measurement of $\langle M_{\rm ap}^2 \rangle(\theta)$ obtained using  our COSMOS galaxy catalogue. The signal has been divided into gravitational (black diamonds) and non-gravitational (red triangles) component. The right panel shows the measurement of $\langle M_{\rm ap}^3\rangle(\theta)$ (black diamonds) and $\langle M_{\rm ap} M_\times^2 \rangle(\theta)$ (red triangles). The statistical noise affecting the E and B modes  is plotted using solid-line error-bars. The total noise on the E modes including sampling variance is shown using dashed-line error-bars. They have been computed using the set of simulations II . The measurement has been compared to a fiducial model obtained using a WMAP7 best-fit cosmology  for the source redshift
distribution of the shear  catalogue (pink solid line).}
\end{figure*}

 The measurement of the two- and three-point correlation functions on the COSMOS field is accomplished by using a tree-code. The tree is built in a way analogous to the one described by Zhang \& Pen (2005) and by Jarvis, Bernstein \& Jain (2004). Each node of the tree is characterised  by the average  ellipticity  $\langle{\bf e}\rangle=\langle e_x\rangle +i \langle e_y \rangle $  over  a region with centre ${\bf x_0}$, and  characteristic size $d$.
 The characteristic size $d$ is defined as the distance from the centre to the most distant galaxy included in the average.
The step from a level to the next one is done by dividing the size $d$ by two and recomputing the average quantities of the two new nodes; the new nodes are called ``children''. At the end of the division process  we obtain a tree which  has a top-level node $(\langle{\bf e}\rangle,{\bf x_0},d)$  computed over the whole image, whereas the  bottom level  has leaves $({\bf e},{\bf x_0},0)$ containing single galaxies; each node is  linked to his children, the ensemble of children from the top node to a leaf is called branch.

We compute the two-point correlation functions  between $\theta_{\rm min}=5~\rm{arcsec}$ and $\theta_{\rm max}=50~{\rm arcmin}$ using logarithmic bins with size  $\delta {\rm ln} \theta$ is $0.05$.  Starting from the top level,  one  must  descend  along the tree to compute the correlation functions between nodes of different branches. In order to decide whenever to continue the descent or to stop and compute the correlation functions between two nodes one must  use a criterion. In our case we use the following criterion:
$
\frac{d_1+d_2}{s}< \delta {\rm ln} \theta
$ where  $d_1$ and $d_2$ are the characteristic size of the nodes  and $s$ is the distance between the centres.  When the condition is verified  we compute the correlation functions between the two nodes and move to other branches. One can see that the smaller  $\delta {\rm ln}  \theta$  is the deeper one needs to descent along the branches to verify the above criterion.  When the criterion is satisfied we compute
\be
\xi_+={\bf e_1 e_2}^{\star}w_1w_2~\;{\rm and}~\;  \xi_-={\bf e_1 e_2} \exp^{-4i\alpha} w_1w_2
\ee
where $\alpha$ is the angle between the x-axis and the line connecting the centres;  ${\bf e_1}$, ${\bf e_2}$ are the average complex ellipticities and $w_1$, $w_2$ are the weights associated to each node, and $\star$ denotes the complex conjugate. We chose this weight  to be equal to the number of the galaxies contained in the  node. Averaging over all the pairs belonging to the same distance bin  we obtain  a measurement of  the two-point  correlations $\xi_+(\theta)$ and $\xi_-(\theta)$.

For the three-point shear statistics  we adopt a similar strategy, but now we have to consider that the correlations are function of three variables.   Using the same notations as Jarvis, Bernstein \& Jain (2004) we call ${\bf s}$  the vector connecting the closest pair characterised by sizes $d_2$, $d_3$. The vector ${\bf t}$ connecting the second closest pair with sizes  $d_2$ and $d_1$, thus  $s<t<|{\bf t}- {\bf s}|$.

We compute the correlation functions for triangles with sides longer than $\theta_{\rm min}=5~\rm{arcsec}$ and smaller than  $\theta_{\rm max}=50~\rm{arcmin}$  using again a logarithmic step  with width  $\delta {\rm ln} \theta=0.1$. 
The criterion for the descent along the tree is similar to the one used for the two-point  shear statistics
\begin{equation}
\frac{d_1+d_2}{s}<\delta  {\rm ln}\theta,~\frac{d_2+d_3}{t}<\delta {\rm ln} \theta,~\frac{d_3+d_1}{|{\bf t}- {\bf s}|}<\delta  {\rm ln}\theta\;.
\end{equation}
When the above conditions are satisfied, we measure the four complex `natural components' as suggested by Schneider \& Lombardi (2003) and by Jarvis, Bernstein \& Jain (2004). They are defined as follows:
\bea
\Gamma_0={\bf e_1e_2e_3}\exp^{-6i\alpha}w_1w_2w_3\;, ~~ \Gamma_1={\bf e^\star_1e_2e_3}\exp^{-2i\alpha}w_1w_2w_3\nonumber\;,~~
\Gamma_2={\bf e_1e^\star_2e_3}\exp^{-2i\alpha}w_1w_2w_3\nonumber\;,~~
\Gamma_3={\bf e_1e_2e^\star_3}\exp^{-2i\alpha}w_1w_2w_3\nonumber
\eea
where $\alpha$ is the angle between ${\bf s}$  and the x-axis.

We show in Figure   \ref{fig:correlations}  the amplitude of the two- and three-point correlations functions measured from the COSMOS galaxy catalogue. The left panel shows the amplitude of the correlation $\xi_+(\theta)$, and the real component of $\xi_-(\theta)$ as a function of the angular scale $\theta$; the top-right (bottom-right) panel shows the measurement of the real (imaginary) part of the four complex `natural components'  $\Gamma_i$  measured on equilateral triangles as a function of the sides length $\theta$.
 
In this paper we are mainly interested in three-point shear statistics and we could use the measurement of the correlation functions $\Gamma_i$ directly to infer cosmological constraints \cite{Scetal05}. However,  it is much more convenient to combine the correlation functions to obtain the third-order moment of the aperture mass  $\langle M_{\rm ap}^3\rangle(\theta)$. In fact, not only the  relation between $\langle M_{\rm ap}^3\rangle(\theta)$  and the bispectrum is much simpler than the one between the correlation functions and the bispectrum, but also, as we pointed out already, the aperture mass statistics   allow one to separate the shear field in E- end B-modes providing a powerful way  to detect   systematics.

By choosing
\be\label{eq:filter}
Q_\theta(\vartheta)=\frac{\vartheta^2}{4\pi\theta^2} \exp{\Big(-\frac{\vartheta^2}{2\theta^2}\Big)}\,,\; U_\theta(\vartheta)=\frac{\vartheta^2}{2\pi\theta^2} \big(1-\frac{\vartheta^2}{4\pi\theta^2}\big)\exp{\Big(-\frac{\vartheta^2}{2\theta^2}\Big)}\,,
\ee
the relations between $\langle M_{\rm ap}^2\rangle(\theta)$ and $\langle
M_{\rm ap}^3\rangle(\theta)$ and two- and three-point correlation
functions are analytic \cite{Peetal03}. 
With this particular choice we obtain for the two-point shear statistics:
\bea\label{eq:2pt_corr}
\langle M^2\rangle(\theta)&\equiv&\langle M_{\rm ap}^2\rangle(\theta)- \langle M_{\rm  \times}^2\rangle(\theta)+2 i \langle M_{\rm ap} M_{\rm \times} \rangle(\theta)=\int \frac{\vartheta d\vartheta}{\theta^2}\xi_-(\vartheta)T_-\big(\frac{\vartheta}{\theta}\Big)\,,\\
\langle MM^\star \rangle(\theta)&\equiv& \langle M_{\rm ap}^2\rangle(\theta)+\langle M_{\rm  \times}^2\rangle(\theta)=\int\frac{\vartheta d\vartheta}{\theta^2} \xi_+(\vartheta)T_+\Big(\frac{\vartheta}{\theta}\Big)\,,
\eea
with
\bea
T_+(x)=\frac{x^4-16x^2+32}{128}\exp\Big(-\frac{x^2}{4}\Big)\;,~~~T_-(x)=\frac{x^4}{128}\exp\Big(-\frac{x^2}{4}\Big)\,.
\eea
In a pure cosmic shear field only the E-mode component $\langle M_{\rm ap}^2 \rangle(\theta)$ is expected to be non zero whereas the B-mode component $\langle M_{\rm \times}^2\rangle(\theta)$ can be used to investigate systematics, and the mixed component is expected to vanish if the non-gravitational modes are parity invariant.  For the three-point shear statistics \cite{Peetal03,Jaetal04,Scetal05} the relations are:
\be
\langle M^3\rangle(\theta)\equiv\langle M_{\rm ap}^3\rangle(\theta)+3i\langle M_{\rm ap}^2M_\times\rangle(\theta) -3\langle M_{\rm ap}M_\times^2\rangle(\theta)-i\langle M_\times^3\rangle(\theta)=6 \int_{s<t<|{\bf t}-{\bf s}|} \frac{\vartheta d\vartheta}{\theta^2}\int\frac{d^2{\bf t}}{2\pi \theta^2}\Gamma_0(s,{\bf t})T_0\Big(\frac{s}{\theta},\frac{{\bf t}}{\theta}\Big)\,,
\ee
\bea
\langle M^2 M^\star\rangle(\theta)&\equiv&\langle M_{\rm ap}^3\rangle (\theta)+i\langle M_{\rm ap}^2M_\times\rangle(\theta)+\langle M_{\rm ap}M_\times^2\rangle(\theta)+i\langle M_\times^3\rangle(\theta)\nonumber\\&=&2\int_{s<t<|{\bf t}-{\bf s}|} \frac{\vartheta d\vartheta}{\theta^2}\int\frac{d^2{\bf t}}{2\pi \theta^2}\Big[\Gamma_1(s,{\bf t})T_1\Big(\frac{s}{\theta},\frac{{\bf t}}{\theta}\Big)+\Gamma_2(s,{\bf t})T_2\Big(\frac{s}{\theta},\frac{{\bf t}}{\theta}\Big)+\Gamma_3(s,{\bf t})T_3\Big(\frac{s}{\theta},\frac{{\bf t}}{\theta}\Big)\Big]\nonumber\,,
\eea
where  
\ba
T_0(s,{\bf t})&=&-\frac{{\bf q}^{\star 2}_1{\bf q}^{\star 2}_2{\bf q}^{\star 2}_3}{24}\exp\Big({-\frac{q_1^2+q_2^2+q_3^2}{2}}\Big)\,,\\
T_1(s,{\bf t})&=&-\Big( \frac{{\bf q}^{2}_1{\bf q}^{\star 2}_2{\bf q}^{\star 2}_3}{24}-\frac{ q^{2}_1{\bf q}^{\star}_2{\bf q}^{\star}_3}{9}+\frac{{\bf q}_1^{\star 2}+2{\bf q}_2^{\star 2}{\bf q}_3^{\star 2}}{27} \Big)\exp\Big({-\frac{q_1^2+q_2^2+q_3^2}{2}}\Big)\,,
\ea
and
\be
{\bf q_1}=({\bf s}+{\bf t})/3s\,,~~~~{\bf q_2}=({\bf t}-2{\bf s})/3s\,,~~~~{\bf q_3}=({\bf s}-2{\bf t})/3s\,,
\ee
and $T_2$, $T_3$ are obtained from $T_1$ by a cyclic rotation of the indices.

In the case of pure cosmic shear signal one expects to measure only the E-mode component $\langle M_{\rm ap}^3 \rangle(\theta)$, whereas the two mixed components $\langle M_{\rm ap}^2M_\times\rangle (\theta)$ and $\langle M_{\rm ap}M_\times^2\rangle(\theta)$ and the B-mode component $\langle M_\times^3\rangle(\theta)$ are expected to vanish. If the observed field contains non-gravitational modes, the mixed component $\langle M_{\rm ap} M_\times^2 \rangle(\theta)$ will not vanish whereas the two other components are still expected to vanish  if the non-gravitational modes are parity invariant. Hence, $\langle M_{\rm ap} M_\times^2 \rangle(\theta)$  can be used to quantify the amplitude of potential residual systematics in the galaxy catalogue.

One notices that the filter in Equation (\ref{eq:filter}) has
 infinite support thus,  in principle, one cannot compute $\langle M_{\rm ap}^2 \rangle(\theta)$ and $\langle M_{\rm ap}^3\rangle(\theta)$ unless the survey has infinite size. 
 However, the filter has an exponential cutoff, thus in practice one can truncate the integration. Simulations show that to compute the
 second- and third-order moments of the aperture mass up to a scale $\theta$ one needs to compute the
 correlation functions up to a scale $\sim 4\theta$ \cite{Jaetal04}. We computed the correlation function up to scales of $50$ arcmin and the second- and third-order moments of the aperture mass up to a scale of 12 arcmin. 
 
The left panel of Figure \ref{fig:map} shows the measured $\langle M_{\rm ap}^2 \rangle(\theta)$ and $\langle M_{\rm \times}^2\rangle(\theta)$ after integration of the two-point correlation functions. The measured cosmological signal has been compared to a fiducial model obtained using a WMAP7 best-fit cosmology \cite{Koetal10} computed using Equation (\ref{eq:map2}) with a redshift distribution given by Equation (\ref{eq:pofz}). The power spectrum in the non-linear regime is computed using Smith et al. (2003) and the transfer function introduced by Eisenstein \& Hu (1998). 

The error-bars for  $\langle M_\times^2 \rangle(\theta)$ include only the noise due to the intrinsic shape of the galaxies. They have been derived using the noise maps computed as described in Section \ref{sec:covariance}.
We assumed that the error-bars on the non-gravitational modes are given only by shape-noise, although this might not be exactly the case as the B-modes might have a power spectrum if the PSF correction has some non-trivial residual pattern or if one accounts for intrinsic alignment effects. Assuming that the uncertainty is only the result of shape noise
would underestimate the total error; nevertheless, as long as the amplitude of $\langle M_\times^2 \rangle(\theta)$ is much smaller than $\langle M_{\rm ap}^2 \rangle(\theta)$ the measurement is not significantly affected by systematics. 
In our case, a linear fit performed using the shape-noise covariance matrix suggests a constant value \mbox{$\langle M_\times^2 \rangle(\theta) = 0.13 \pm 1.86 \times 10^{-7}$}, which is  much smaller that the estimated cosmological shear signal\footnote{Note that Schrabback et al. (2010) used a different filter
function for their E-/B-mode decomposition using $\langle 
M^2_\mathrm{ap}\rangle (\theta)$, for which they estimated errors from bootstrap 
resamples,
leading to slight differences in both the signal and errors compared
to our results.}.
For $\langle M_{\rm ap}^2 \rangle(\theta)$ we show both the shape (solid error-bars)  and the total noise (dashed error-bars).
The right panel of Figure \ref{fig:map} shows the measurement of the third-order moment of the aperture mass components $\langle M_{\rm ap}^3\rangle(\theta)$, $\langle M_{\rm ap} M_\times^2 \rangle(\theta)$  and the fiducial model computed for a WMAP7 best-fit cosmology using Equation (\ref{eq:map3}) and Equation (\ref{eq:pofz}). At small scales the amplitude of $\langle M_{\rm ap}^3\rangle(\theta)$ is much larger than the one of the non-gravitational component $\langle M_{\rm ap} M_\times^2 \rangle(\theta)$. The amplitude of $\langle M_{\rm ap} M_\times^2 \rangle(\theta)$ is consistent with zero, whereas the amplitude of the cosmic shear signal is in good agreement with the one suggested by the WMAP7 cosmology. At large scales 
the weak lensing signal becomes very small and thus its amplitude is comparable with the one of the mixed component, but still the amplitude   of the mixed component is consistent with zero.
The error-bars  for $\langle M_{\rm ap} M_\times^2 \rangle(\theta)$ include only shape noise, whereas for  $\langle M_{\rm ap}^3\rangle(\theta)$ we showed both shape (solid error-bars) and total noise (dashed error-bars).  As we will discuss in more detail in Section \ref{sec:cosmodep}, the sampling variance is the dominant source of noise for our set of data.

\subsection{Testing for systematics}\label{sec:systematics}
\begin{figure*}
\begin{tabular}{cc}
\psfig{figure=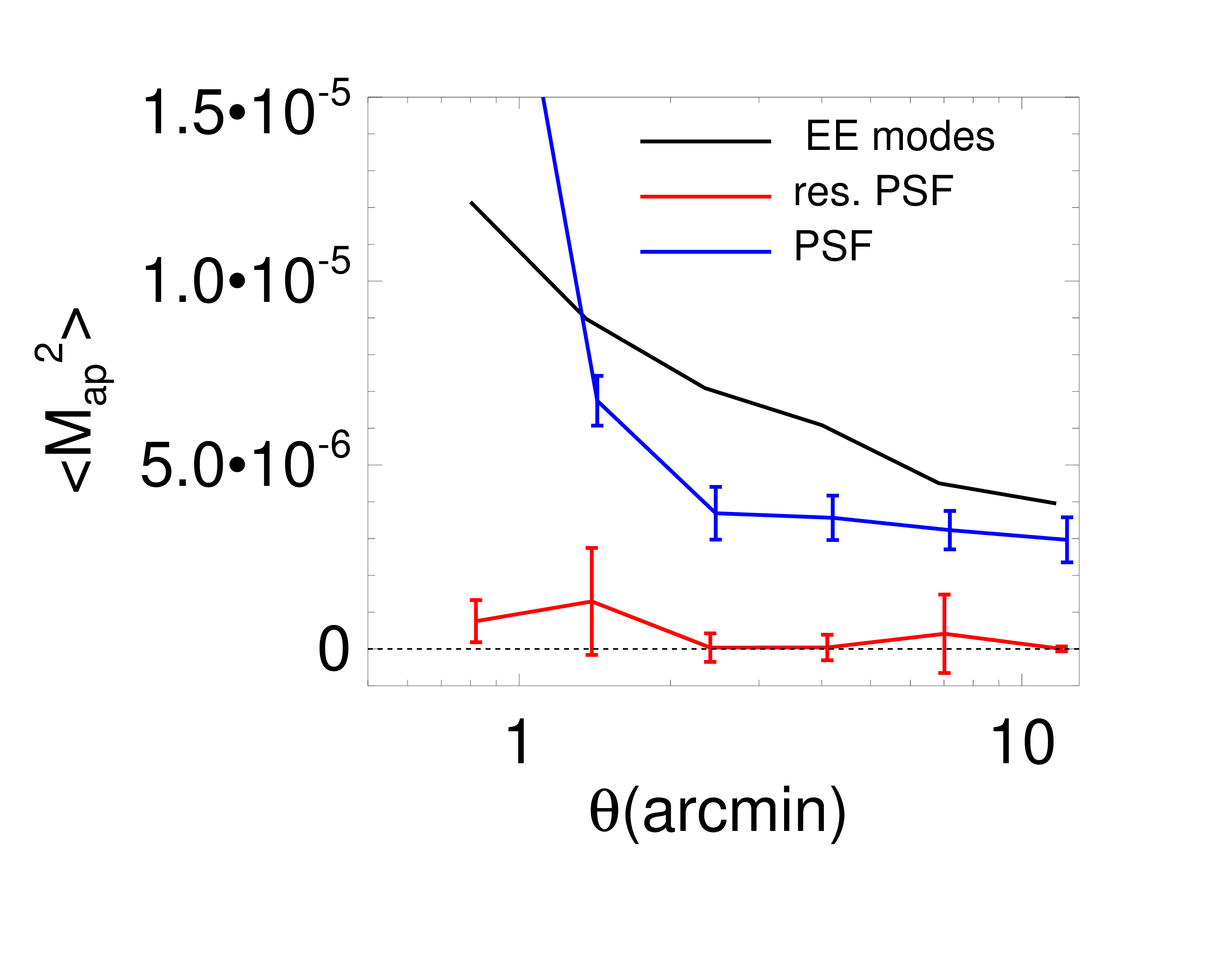,width=.45\textwidth}&\psfig{figure=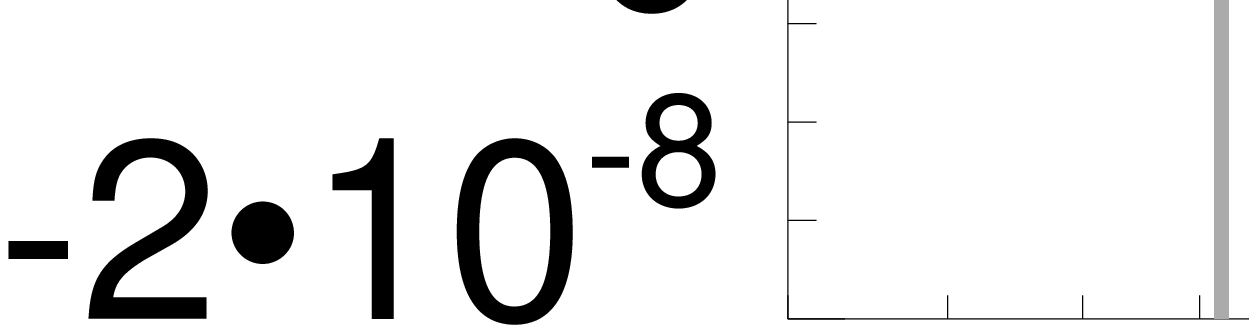,width=.45\textwidth}
\end{tabular}
	\caption{\label{fig:systematics}Left panel: the amplitude of the cosmological signal $\langle M_{\rm ap}^2\rangle(\theta)$ (black solid line) is compared with the measurement of $\langle M_{\rm SYS}^2\rangle(\theta)$ (red solid line) defined by Equation (\ref{eq:sys_2pt}) which is expected to be non-zero for non-optimal PSF correction. For comparison we also show the amplitude of the aperture mass variance measured on the uncorrected stars, $\langle M^2_{ss} \rangle (\theta)$ (blue solid line). 
	Right panel: the amplitude of the measured cosmological signal $\langle M_{\rm ap}^3\rangle(\theta)$ (black solid line) is compared with the measurement of $\langle M_{1,\rm SYS}^3\rangle(\theta)$ and $\langle M_{2,{\rm SYS}}^3\rangle(\theta)$ (gray and red solid lines) defined by Equations (\ref{eq:sys_3pt_1}) and (\ref{eq:sys_3pt_2}). We plot for comparison the amplitude  of the statistics $\langle M_{ sss}^3\rangle(\theta)$ measured on the uncorrected stars. The error-bars on each statistic include only shape-noise.}
\end{figure*}
For a pure cosmic shear field one expects only E-type correlations between the observed ellipticity of galaxies. This means that whenever one finds B-modes it is likely that the observed alignment between galaxies is produced by something other than cosmic shear, for example PSF correction residuals. 
Therefore, the division of the two- and three-point shear statistics into E, B and mixed components shown in the previous section can be used to investigate the presence of systematic errors.
 
 However, one cannot expect the PSF residual to affect the E, B and mixed component in the same way. A more direct way to quantify directly the correlation between PSF and measured galaxy ellipticity is given by \cite{Baetal03,Waetal05}: 

 \be
 \xi_{\rm SYS}(\theta)= \frac{\xi_{gs}^2}{\xi_{ss}}(\theta ) 
 \ee
where $\xi_{gs}$ is the correlation between the ellipticity of uncorrected stars, i.e. the PSF ellipticity, and the ellipticity of the PSF corrected galaxies.  Here, $\xi_{ss}$ is the correlation between the ellipticities of the uncorrected stars. $\xi_{\rm SYS}(\theta)$ is normalised such  that its amplitude can be directly compared with the amplitude of the cosmological signal, i.e. the correlation function $\xi(\theta)$ measured between pairs of galaxies. In fact, in the case of a perfect PSF correction the correlation between stars and galaxies would be zero so $\xi_{\rm SYS}(\theta)$ would also be zero. On the other hand, if the PSF and the estimated shear are perfectly correlated the amplitude of $\xi_{\rm SYS}(\theta)$ is about the same as $\xi(\theta)$.

In a more general way, one can define cross-correlation two-point statistics for other types of filters such as the aperture mass. Integrating $\xi_{gs}(\theta)$ and $\xi_{ss}(\theta)$ one obtains the aperture masses ${\langle M^2_{gs}\rangle(\theta)}$ and ${\langle M^2_{ss}\rangle(\theta)}$ and can define:
\be \label{eq:sys_2pt}
 \langle M_{\rm SYS}^2\rangle(\theta)=\frac{\langle M^2_{gs}\rangle^2}{\langle M^2_{ss}\rangle}(\theta) 
\ee 
which is also normalised so that it can be compared with $\langle M_{\rm ap}^2\rangle(\theta)$. %One notices that we do not divide $\langle M_{\rm SYS}^2\rangle(\theta)$ into gravitational and non-gravitational componenents. 

For the three-point shear statistics we define the correlation functions: $\xi_{gss} (\theta_1,\theta_2,\theta_3)$ between two stars and a galaxy, $\xi_{ggs} (\theta_1,\theta_2,\theta_3)$ between two galaxies and a star, $\xi_{sss} (\theta_1,\theta_2,\theta_3)$ between three stars; here $\theta_1$, $\theta_2$, $\theta_3$ denote the distances between the vertices of the triangle. Using these correlation functions we define the third-order aperture mass statistics $\langle M^3_{gss}\rangle(\theta)$, $\langle M^3_{ggs}\rangle(\theta)$ and $\langle M^3_{sss}\rangle(\theta)$.  Finally we define:
\ba
 \langle M_{1,{\rm SYS}}^3\rangle(\theta)&=& \frac{\langle M^3_{gss}\rangle^3}{\langle M^2_{ss}\rangle^3}(\theta)\label{eq:sys_3pt_1}\,,\\
 \langle M_{2,{\rm SYS}}^3\rangle(\theta)&=& \frac{\langle M^3_{ggs}\rangle ^{3/2}}{\langle M^2_{ss} \rangle^{3/4}} (\theta) \label{eq:sys_3pt_2}\,,
 \ea
 which are normalised so they can be directly compared with $\langle M_{\rm ap}^3\rangle(\theta)$. 
 \begin{figure*}
\begin{tabular}{cc}
\psfig{figure=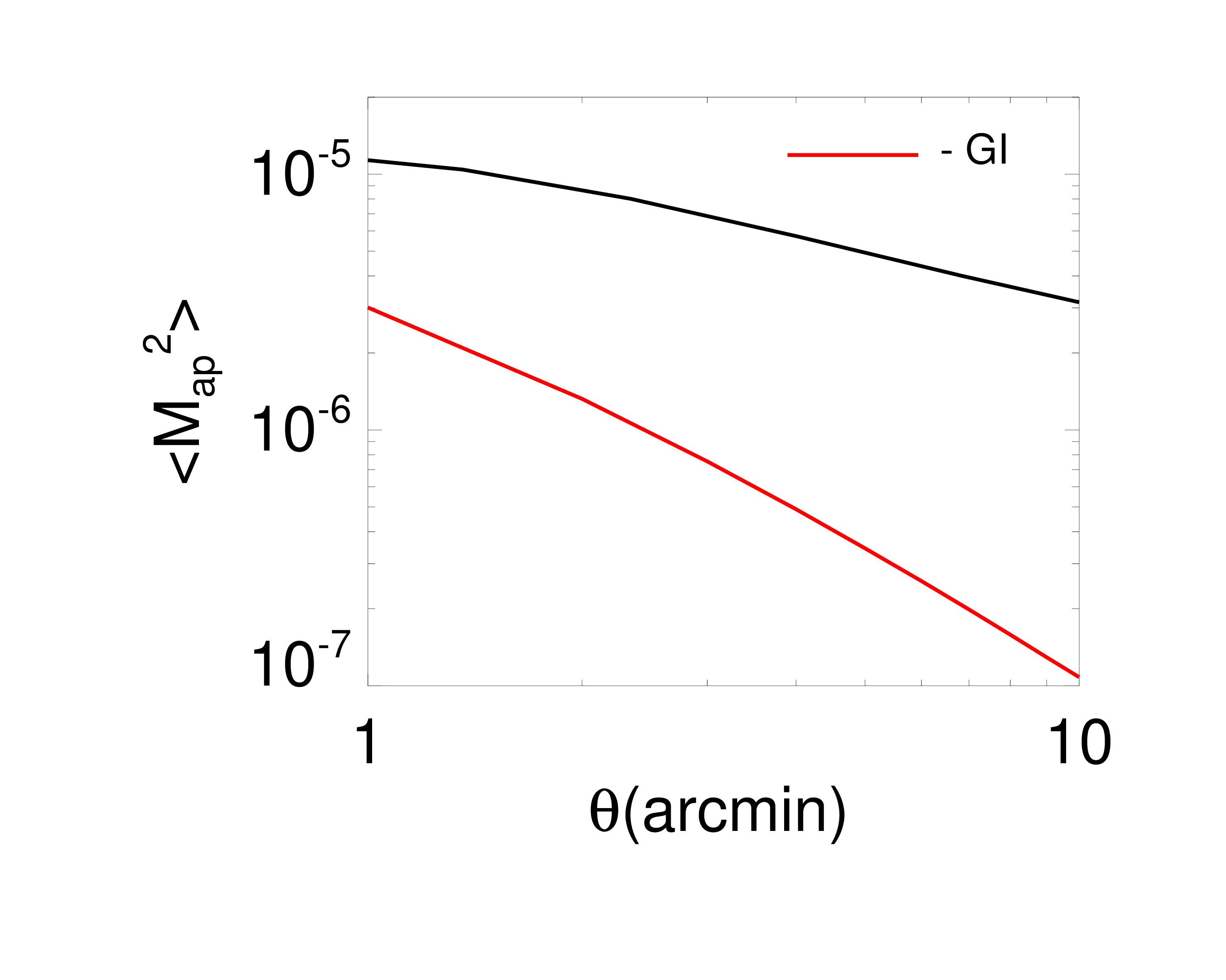,width=.45\textwidth}&\psfig{figure=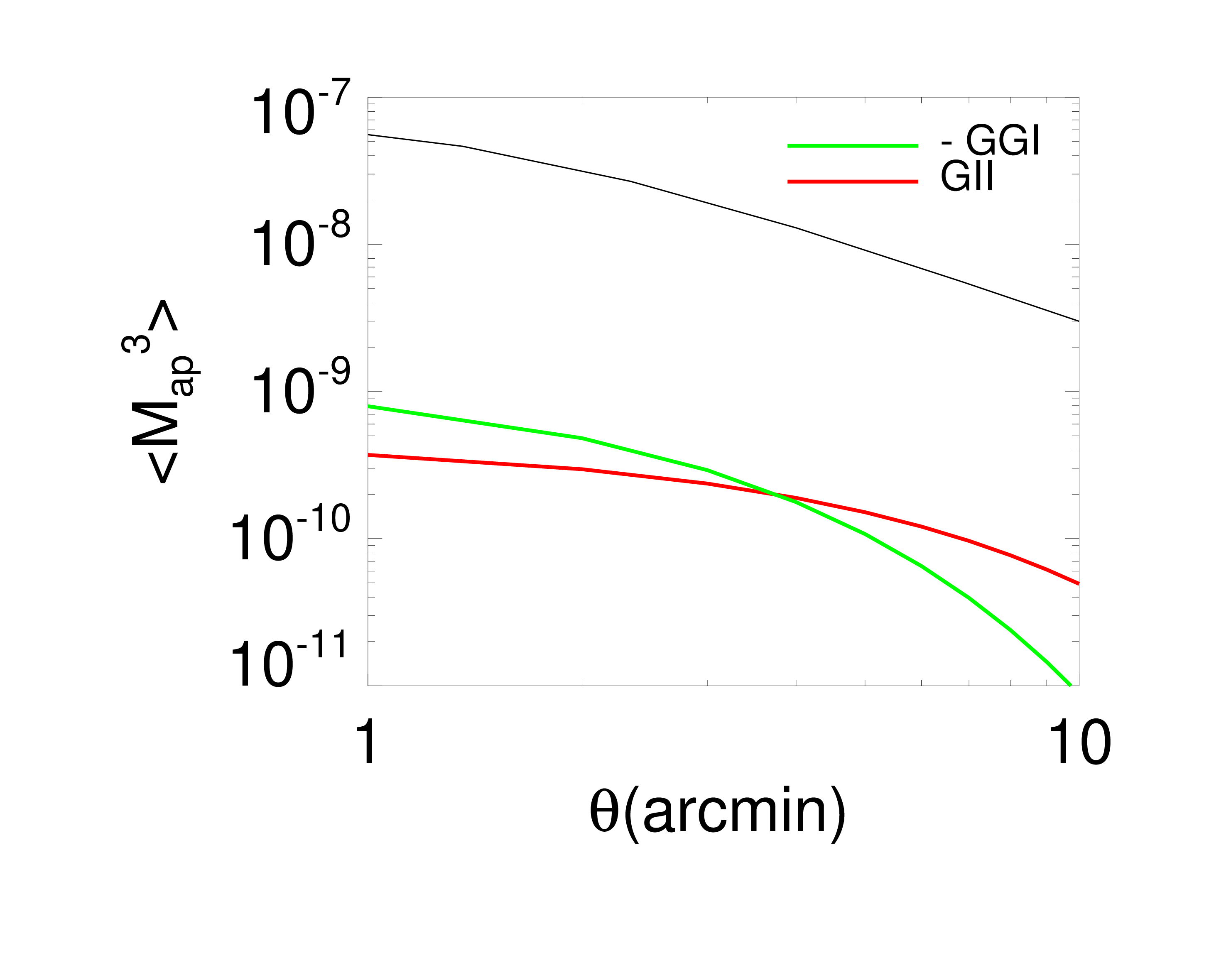,width=.45\textwidth}
\end{tabular}
\caption{\label{fig:intrinsic} Left panel: the amplitude of the cosmological signal $\langle M_{\rm ap}^2\rangle(\theta) $ expected for a survey with the same depth as COSMOS (black line) is compared with the amplitude of the expected $|{\rm GI}|$ (red line) for a survey with the same redshift distribution. The GI model has been computed using Equation (\ref{eq:GI}) with the values $A_{\rm GI}=-1.29 \times 10^{-7}~ h~ {\rm Mpc} ^{-1}$ and $\theta_{\rm GI}=0.93$ arcmin which has been quoted by Heymans et al. (2006b) as the best fit-values for a model for a Universe composed of about $30\%$ elliptical and $70\%$ spirals. Right panel: the amplitude of the expected cosmological signal $\langle M_{\rm ap}^3\rangle(\theta)$ (black line) is compared with  GII (red line) and $|{\rm GGI}|$ (green line). The values for GII and GGI have been obtained by using the following values: $A_{\rm GII}= 0.07 \times 10^{-7} ~h ~{\rm Mpc} ^{-1}$ and $\theta_{\rm GGI} =3.74$ arcmin, and $A_{\rm GII}=-0.22 \times 10^{-7} ~h ~{\rm Mpc} ^{-1}$ and $\theta_{\rm GGI}= 1.61$ arcmin, which are the best-fit models presented by Semboloni et al. (2008) for the same galaxy population used by Heymans et al. (2006b).}
\end{figure*}

The left panel of Figure \ref{fig:systematics} shows the amplitude of the cosmic shear signal, the amplitude of $\langle M_{ss}^2\rangle (\theta)$ and the amplitude of $\langle M_{\rm SYS}^2\rangle(\theta)$. The amplitude of $\langle M_{ss}^2\rangle(\theta)$ is as high as the cosmic shear signal, whereas the amplitude of $\langle M_{\rm SYS}^2\rangle(\theta)$ is much smaller, meaning that the PSF is well corrected.
The error-bars for $\langle M_{ss}^2\rangle(\theta)$ and $\langle M_{\rm SYS}^2\rangle(\theta)$ have been computed assuming Gaussian shape-noise both for galaxies and stars.
The right panel of Figure \ref{fig:systematics} shows the amplitude of $\langle M_{\rm ap}^3\rangle(\theta)$, $\langle M^3_{sss}\rangle(\theta)$, $\langle M_{1,{\rm SYS}}^3\rangle(\theta)$ and $\langle M_{2,{\rm SYS}}^3\rangle(\theta)$. It is very interesting to notice that the raw PSF does not show significant three-point correlation at least at small scales where the cosmological signal carries the most of the information. The PSF correction is very good at small scales thus, the cross correlation is always much smaller than the cosmological signal. 

One notices, that the definition of $\langle M_{1,{\rm SYS}}^3\rangle(\theta)$ and $\langle M_{2,{\rm SYS}}^3\rangle(\theta)$ is not unique. Indeed, one could have defined them by normalising $\langle M^3_{ggs}\rangle(\theta)$ and $\langle M^3_{gss}\rangle(\theta)$ by a suitable power of $\langle M^3_{sss}\rangle(\theta)$ instead than a power of $\langle M^2_{ss}\rangle(\theta)$. 
 We find that the estimators $\langle M_{1,{\rm SYS}}^3\rangle(\theta)$ and $\langle M_{2,{\rm SYS}}^3\rangle(\theta)$ defined in that way also suggest very low PSF residuals. Nevertheless, they are much more noisy than the ones shown in the plots of Figure \ref{fig:systematics} especially at small scales where the triplets of stars are very few.

\subsection{Intrinsic ellipticity correlations}
 If galaxies are randomly oriented one can  statistically interpret the observed correlation between galaxies as pure cosmic shear effect. However,  galaxies belonging to clusters are affected by the local
gravitational tidal forces. This effect creates an intrinsic alignment between galaxies which are physically close, so strictly speaking the hypothesis that galaxies are randomly oriented is not valid. Hirata \& Seljak (2004) pointed out that the local tidal field affects the measure of two-point shear statistics through two terms: one of intrinsic alignments of physically associated galaxies and the other generally called shape-shear, which describes the correlation between the intrinsic shape of foreground galaxies with their local density field which is causing the shear of background galaxies.

In the past years intrinsic alignment has been indicated as one of the potential sources of systematics affecting the measurement of cosmic shear statistics. Using $N$-body simulations, Heymans et al. (2006b) found that for shallow surveys the intrinsic alignment can significantly affect the measured two-point shear statistics. This result agrees with the one that Mandelbaum et al. (2006) derived  using galaxies from the SDSS spectroscopic sample.  Semboloni et al. (2008) showed by using $N$-body simulations that the intrinsic alignment of galaxies can dominate the cosmological signal for three-point cosmic shear signal.  However, according to these results the amplitude of the intrinsic alignment strongly depends on the projections and  should not substantially bias the cosmic shear statistics measured on deep surveys such as COSMOS. 

The shape-shear term couples pairs of galaxies which are at different redshifts and it can in principle significantly affect the measurement of the two- and three-point cosmic shear statistics even for deep surveys.
Heymans et al. (2006b) found that the correlation between the cosmic shear $\gammag$ of a source at redshift $z_s$ and the intrinsic shape $\eg$ of a foreground galaxy at redshift $z_l$ close to the lensing overdensity can be modelled as:
\be\label{eq:GI}
\langle \gammag(z_s) \eg(z_l)\rangle (\theta) =\frac{f_k(z_l) f_k(z_s-z_l)}{f_k(z_s)} \frac{A_{\rm GI}}{\theta+\theta_{\rm GI}}
\ee
where $A_{\rm GI}$ and $\theta_{\rm GI}$ are two parameters which depend on galaxy type.
Semboloni et al. (2008) found using the same set of simulations that the shear-shape correlation affects the measurement of $\langle M_{\rm ap}^3\rangle(\theta)$ through two terms that can be modelled by:
\ba
 {\rm GGI}(z_{s_1},z_{s_2},z_l,\theta)&=&\frac{f_k(z_{s_1}-z_l)}{f_k(z_{s_1})}\frac{f_k(z_{s_2}-z_l)}{f_k(z_{s_2})} A_{\rm GGI}e^{(-\theta/\theta_{\rm GGI})}\label{eq:GGI}\\
 {\rm GII}(z_{s},z_l,\theta)&=&\frac{f_k(z_s-z_l)}{f_k(z_s)}A_{\rm GII} e^{ (-\theta/\theta_{\rm GII})}\label{eq:GII}
 \ea
where the second equation uses the condition that two foreground galaxies that are close are both at the same redshift.
 Once again $A_{\rm GGI}$, $A_{\rm GII}$, $\theta_{\rm GGI}$ and $\theta_{\rm GII}$ are parameters which depend upon the characteristics of the galaxies. 
Among the values of the parameters given by Heymans et al. (2006b) and Semboloni et al. (2008) we use the best-fit values as given for a mixed galaxy population to estimate the amplitude of the shear-shape coupling for the COSMOS survey; the mixed model is composed of about $70\%$ spirals and $30\%$ ellipticals and should be considered the most realistic among the ones analysed in those papers. Using the redshift distribution found in Section \ref{sec:data} we integrate Equations (\ref{eq:GI}), (\ref{eq:GGI}), and (\ref{eq:GII}) along the line-of-sight. 

The left panel of Figure \ref{fig:intrinsic} shows the amplitude of the expected GI component and the amplitude of the expected cosmological signal $\langle M_{\rm ap}^2\rangle(\theta)$ for the same redshift distribution. The amplitude of the cosmic shear signal is always more than 20 times larger than the expected GI contribution. The right panel of Figure \ref{fig:intrinsic} shows the expected amplitude of the two components GGI and GII and the expected cosmic shear signal $\langle M_{\rm ap}^3\rangle(\theta)$. As pointed out by Semboloni et al. (2008) the two components have different signs and the net effect depends on the redshift distribution. It turns out that for a deep survey such as COSMOS the contribution of the two terms is not only small but, the two terms almost cancel out so that the total effect from the shear-shape coupling is expected to have basically no impact on our measurement.

Assuming that the empirical models we used to estimate the shear-shape are correct, the coupling is very likely not a source of significant systematic error for the measurement of the two- and three-point cosmic shear statistics. Nevertheless, we would like to point out that the effect of intrinsic alignment on two- and especially on three-point shear statistics is still poorly known and generally based on approximations and toy-models which may be too simplistic to describe reality. Much more work is needed to rigorously quantify the effect of intrinsic alignment on cosmic shear statistics. 
\section{Covariance matrix estimation}\label{sec:covariance}

The covariance matrix ${\mathcal C}$ can be naturally divided into three components \cite{Scetal02}:
\be\label{eq:covar}
{\mathcal C}={\mathcal C}_{ss}+{\mathcal C}_{sn}+{\mathcal C}_{nn}\;,
\ee
where ${\mathcal C}_{ss}$ is the pure sampling variance term, ${\mathcal C}_{nn}$ is the statistical noise, which we assume is generated only by the intrinsic ellipticity of the sources, ${\mathcal C}_{sn}$ which is a cross term generated by the correlation between cosmological signal and statistical noise.

To estimate the covariance matrix for the two-point shear statistics one can use analytical approximations \cite{Scetal02,Joetal08}. However, analytical approximations assume that the field of matter fluctuations follows a Gaussian statistic; this causes the covariance matrix to be underestimated at small angular scales \cite{Seetal07,TaJa09,Pietal09}. The third-order shear moments are generated by the non-linear evolution of the density matter field, i.e. when the Gaussian approximation is not valid, so it is even less desirable to use the Gaussian approximation to compute the associated covariance matrix. For this reason we compute the covariance matrix using ray-tracing from the Millennium Simulation (i.e. the set simulations II). We divide each of the 32 fields into 9 subfields having the same area as COSMOS, thus obtaining 288 quasi-independent lines-of-sight.  For each subfield we generate shear catalogues  with the same number of galaxies, the same redshift distribution and the same masks as the COSMOS catalogue; finally we need to add shape-noise to each simulated galaxy.  To do so we need to determine the ellipticity dispersion $\sigma_e=\sqrt{\sigma_{e,1}^2+\sigma_{e,2}^2}$ of our catalogue. In order to investigate if we need to vary $\sigma_e$ with redshift, we divide the catalogue in magnitude and size bins and inspect the variance in each bin. Note that the eventual change  of $\sigma_e$ is not necessarily the result of galaxy evolution, but it is mainly due to the fact that the shear estimation is more affected by noise for faint and small objects.  

 We find that the dispersion depends on the magnitude; for bright objects the dispersion is $\sigma_{e}=0.39$, for faint objects $\sigma_{e}=0.49$. However, this is only a minor effect and should barely change the value of $\sigma_e$ as a function of the redshift (the very bright galaxies have a dispersion $20\%$ smaller than the faint ones).  Thus, we decide to use a  dispersion $\sigma_e=0.44$ average of the whole sample,  overestimating the noise at low redshifts. This should not have a significant effect as we expect the sampling variance to be the main source of the noise. Furthermore, we know that most of the contribution to the weak lensing signal comes from sources at high redshift. Once we added the noise to the shear catalogues, we compute the two- and three-point correlation functions on the 288 lines-of-sight and integrate to obtain  $\langle M_{\rm ap}^2\rangle(\theta)$ and $\langle M_{\rm ap}^3\rangle(\theta)$.  
 
Finally, we compute the covariance between two measurements $x_i$ and $x_j$ (where $x_i$ can be either the measurement of $\langle M_{\rm ap}^2\rangle(\theta)$ or $\langle M_{\rm ap}^3\rangle(\theta)$ at a given angular scale) as:
\be
{\mathcal C } (x_i,x_j)=\frac{1}{n_{\rm maps}} \sum_{n=1}^{n_{\rm maps}} (x_{i,n}-\bar{x_i})(x_{j,n}-\bar{x_j})
\ee 
where $\bar{x_i}$ is the measurement of the quantity $x_i$ averaged over the 288 lines-of-sight.
\begin{figure*}
\begin{tabular}{lll}
\psfig{figure=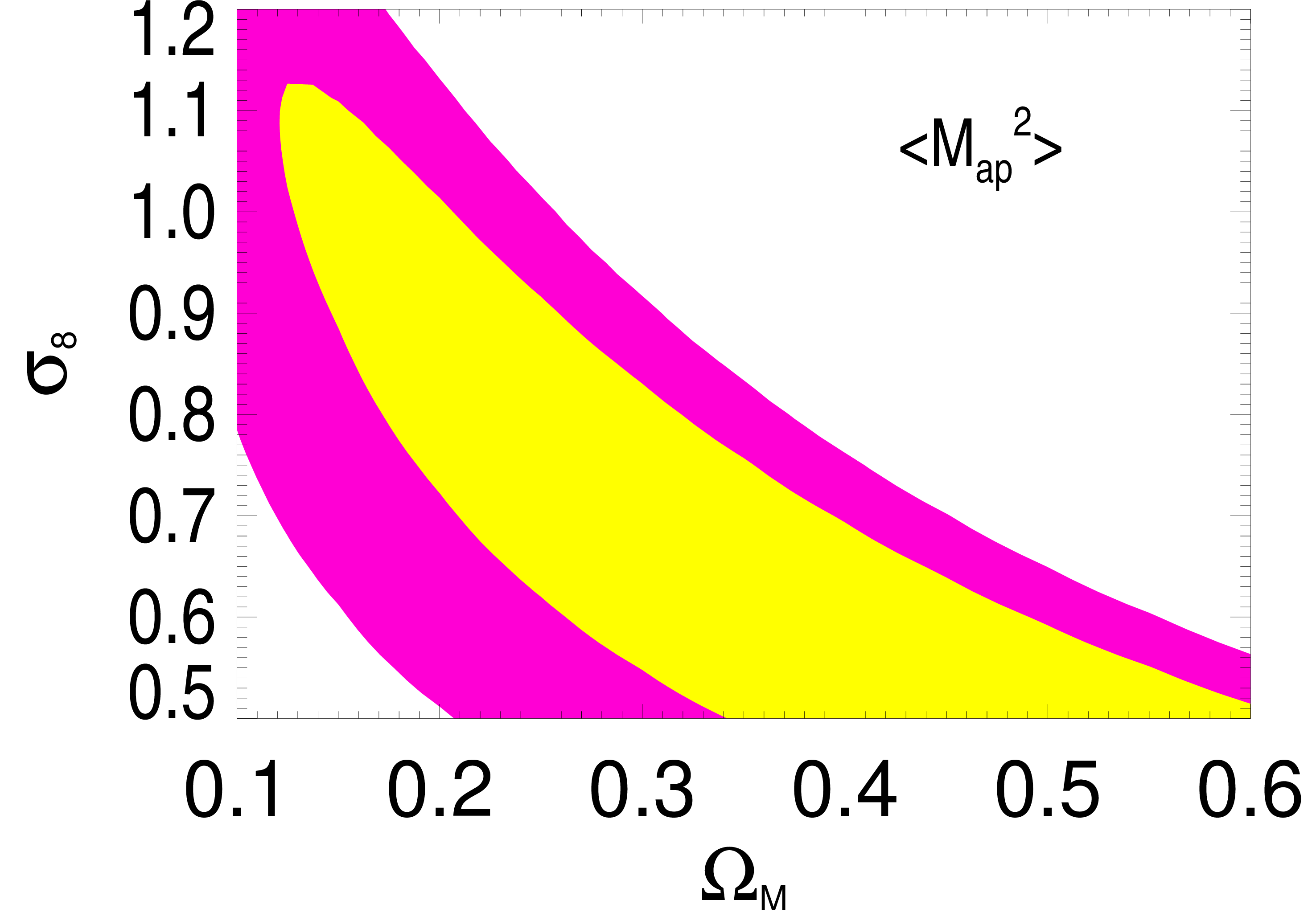,width=.31\textwidth}&\psfig{figure=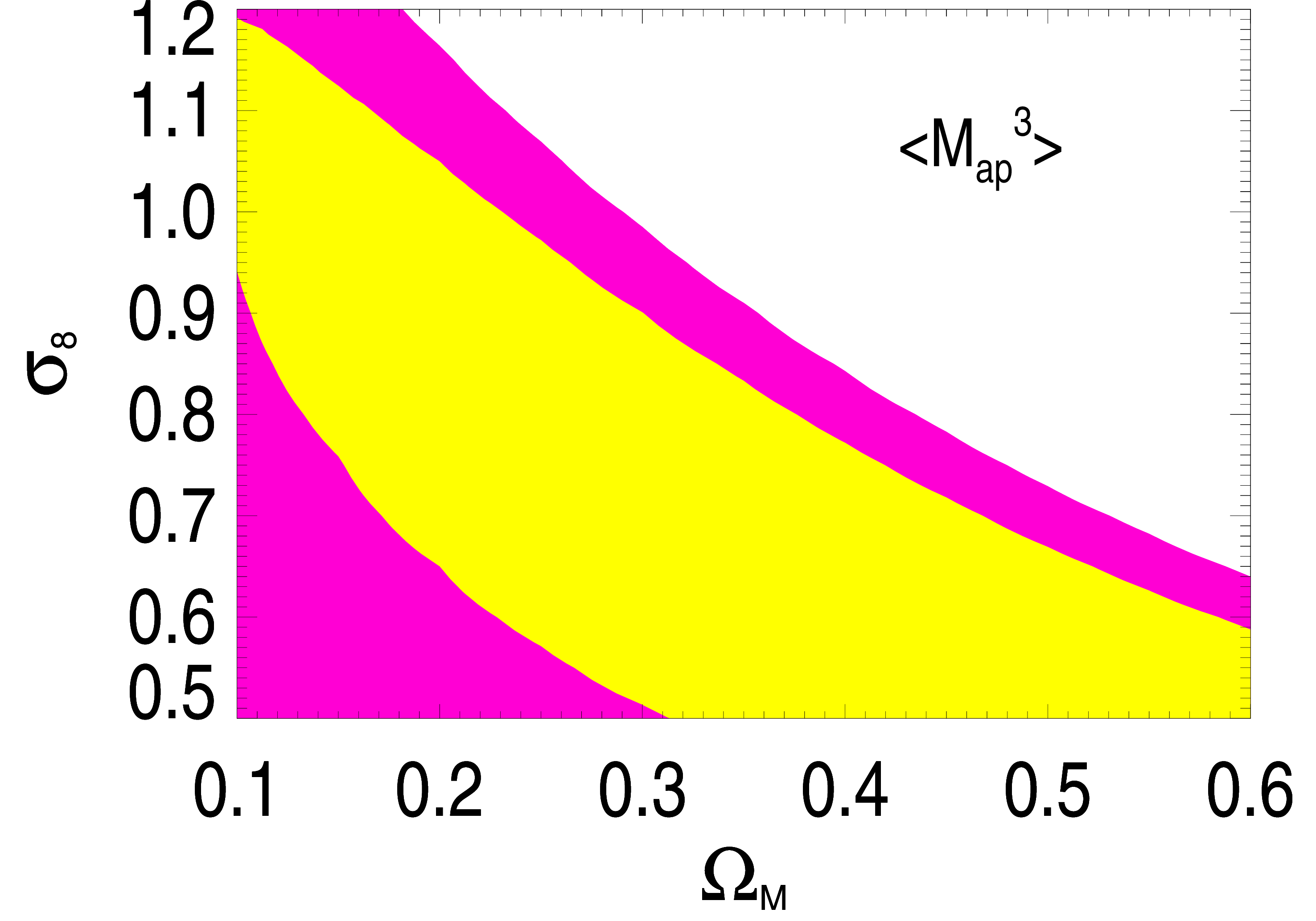,width=.31\textwidth}&\psfig{figure=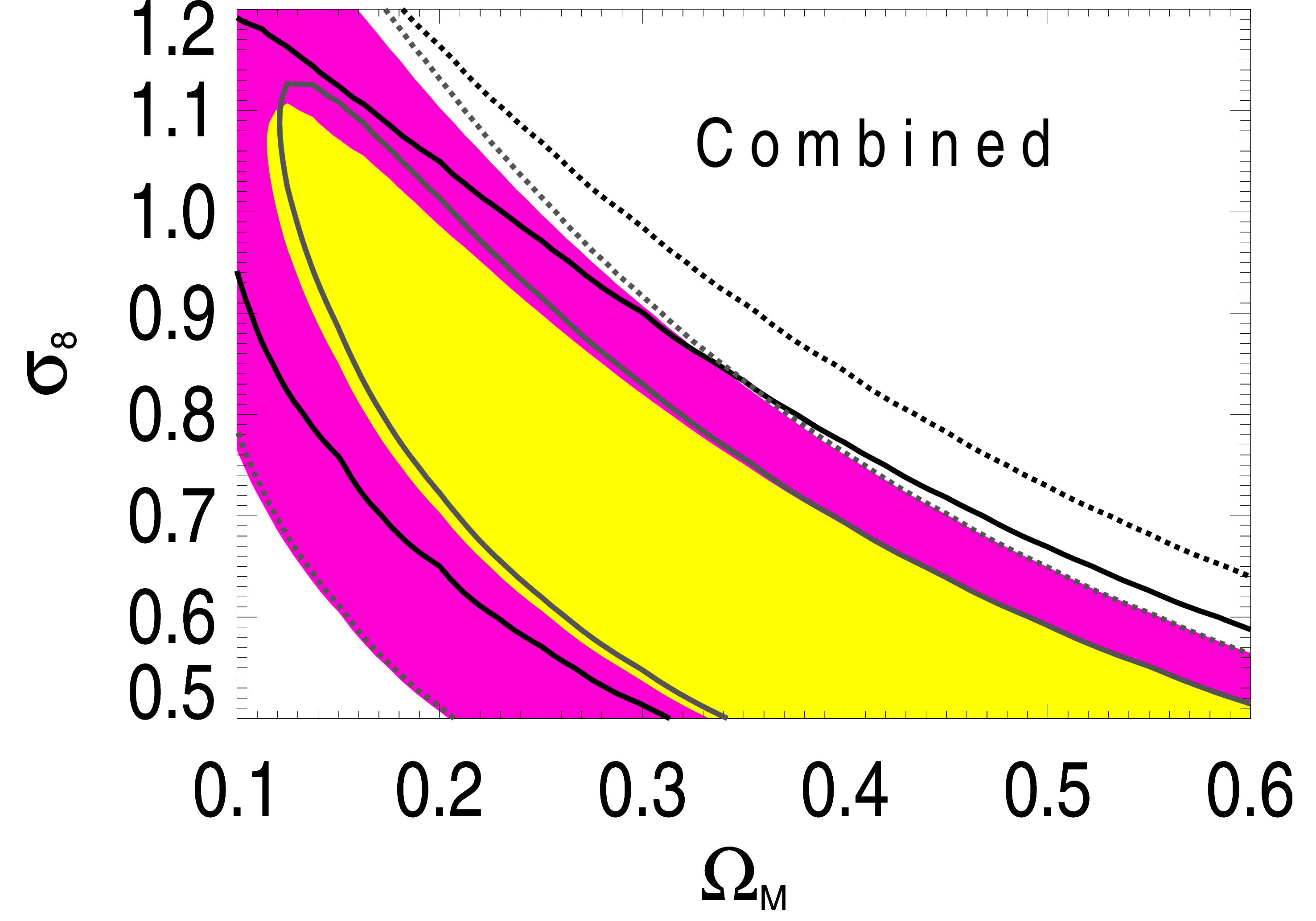,width=.31\textwidth}
\end{tabular}
\caption{\label{fig:results} Left panel: probability distribution for the parameters $\Omega_m$ and $\sigma_8$ obtained preforming a likelihood analysis of the measurement of $\langle M_{\rm ap}^2 \rangle(\theta)$. The two different colours represents the regions where the confidence level is higher than $68\%$ and $95\%$. Middle panel: same as left panel obtained using the measurement of $\langle M_{\rm ap}^3 \rangle(\theta)$. Right panel: probability distribution for the parameters $\Omega_m$ and $\sigma_8$ resulting by a likelihood analysis in which we used the measurement of $\langle M_{\rm ap}^2 \rangle(\theta)$ and $\langle M_{\rm ap}^3\rangle(\theta)$ together. The solid (dashed) lines represent the $68\%$ ($95\%$) level of confidence for $\langle M_{\rm ap}^2 \rangle(\theta)$ and $\langle M_{\rm ap}^3\rangle(\theta)$ used separately as in the left and middle panels. }
\end{figure*}

\section{Cosmological constraints}\label{sec:constraints}

\subsection{Likelihood analysis} 
The measure of two- and three-point shear statistics can be used to infer cosmological constraints on ${\rm \Lambda CDM}$ models. This is done through a likelihood analysis which establishes the posterior probability distribution $P({\bf p}|\xg)$ for the ensemble of parameters ${\bf p}$:
\be\label{eq:posterior}
P({\bf p} |\xg)\propto {\mathcal L} (\xg |{\bf p}) P({\bf p} )\,,
\ee
where $P({\bf p})$ is the prior probability distribution for the ensemble of parameters and $\xg$ is the vector of $n$ measurements. 
We will assume that the likelihood  ${\mathcal L} ({\bf p}|\xg)$ is Gaussian:
\be\label{eq:likelihood}
{\mathcal L} (\xg |{\bf p})\propto \exp\Big(-\frac{1}{2} (\xg-\xg_{m}({\bf p}))^t {\mathcal C}^{-1} (\xg-\xg_{m}({\bf p}))\Big)\,,
\ee
although this is only an approximation, whose limits have been discussed in Hartlap et al. (2009). The vector $\xg$ contains the measurements of $\langle M_{\rm ap}^2 \rangle(\theta)$ and $\langle M_{\rm ap}^3\rangle(\theta)$ presented in Section \ref{sec:measure}, $\xg_m(\thetag)$ is the vector of the models of $\langle M_{\rm ap}^2 \rangle(\theta)$ built using Equation (\ref{eq:map2}) and $\langle M_{\rm ap}^3\rangle(\theta)$ built using Equation (\ref{eq:map3}); ${\mathcal C}^{-1}$ is the inverse of the covariance matrix computed in Section \ref{sec:covariance}. Before inverting it we multiply by the correction factor:
 \be
 \alpha=\frac{(n_{\rm maps}-1)}{(n_{\rm maps}-1)-n-1}\,,
 \ee
which allows one to obtain an unbiased estimation of the inverse of the covariance matrix for a given number of simulations $n_{\rm maps}$ and a number of measurements $n$ \cite{Haetal07}. 

As the COSMOS field covers a small area we do not expect to have strong constraints if we allow a large number of cosmological parameters to vary. Hence, we decide to vary only the  matter density $\Omega_m$ and the power spectrum normalisation  $\sigma_8$ to which the cosmic shear statistics are particularly sensitive. We assume uniform priors for both parameters; we vary $\Omega_m$ between $[0.1,1.]$ and $\sigma_8$ between $[0.5, 1.2]$.  Moreover, we found in Section \ref{sec:data} that the redshift uncertainties can be included in the likelihood analysis by marginalising over the parameter $z_0$ between $[0.443,0.574]$, keeping the other two redshift parameters fixed: $\alpha=1.29$ and $\beta=0.95$.   This interval  in $z_0$  represents the $\pm 3 \sigma$ width of the Gaussian distribution of $p_(z_0)$ found in Section \ref{sec:data}.  We fix all the other cosmological parameters to the best-fit cosmology of WMAP7. 

The left and middle panel of Figure \ref{fig:results} show the likelihood contours in the $[\Omega_m,\sigma_8]$ space, marginalised over the $z_0$ parameter, obtained using $\langle M_{\rm ap}^2 \rangle(\theta)$ and $\langle M_{\rm ap}^3\rangle(\theta)$ respectively; the right panel shows the likelihood contours obtained by combining the measurements of both statistics.
For each of the posterior probability distributions in the $[\Omega_m,\sigma_8]$ space after marginalisation over the redshift parameter $z_0$ we find the local maxima and fit them using a relation $\sigma_8 \Big(\frac{\Omega_m}{0.30}\Big)^X=Y$ where $X$ and $Y$ are free parameters. This allows us to fit  the degeneracy direction between $\Omega_m$ the $\sigma_8$  fairly well obtaining the following results:

\ba\label{eq:results}
&\sigma_8 \Big(\frac{\Omega_m}{0.30}\Big)^{0.63}=0.70^{+0.11}_{-0.14}&{\rm \langle M_{\rm ap}^2\rangle(\theta) }\;,\\
&\sigma_8 \Big(\frac{\Omega_m}{0.30}\Big)^{0.49}=0.78^{+0.11}_{-0.26}&{\rm \langle M_{\rm ap}^3\rangle (\theta)}\;,\\
&\sigma_8 \Big(\frac{\Omega_m}{0.30}\Big)^{0.63}=0.69^{+0.11}_{-0.14}& {\rm \langle M_{\rm ap}^2\rangle (\theta)~ and~ \langle M_{\rm ap}^3\rangle (\theta)}\;,
\ea
where the $1\sigma$ error-bars correspond to the upper and lower limit of the $68\%$ confidence regions.

As expected, the constraints on the cosmological parameters, especially the ones obtained using $\langle M_{\rm ap}^3\rangle(\theta)$ are broad as the COSMOS field is small, and the cosmic variance is large. Moreover, the measurement of $\langle M_{\rm ap}^3\rangle(\theta)$ does not improve the accuracy of the cosmological constraints when used together with the measurement of $\langle M_{\rm ap}^2 \rangle(\theta)$. This is not surprising, as Vafaei et al. (2010) pointed out that the geometry of a narrow and deep field such as COSMOS is not optimal for the measurement of three-point shear statistics. According to their results wide and moderately shallow surveys would be more suitable and would allow one to partially break the degeneracy between  $\Omega_m$ and $\sigma_8$ by using two- and three-point shear statistics. 
 \begin{figure*}
\begin{tabular}{l}
\psfig{figure=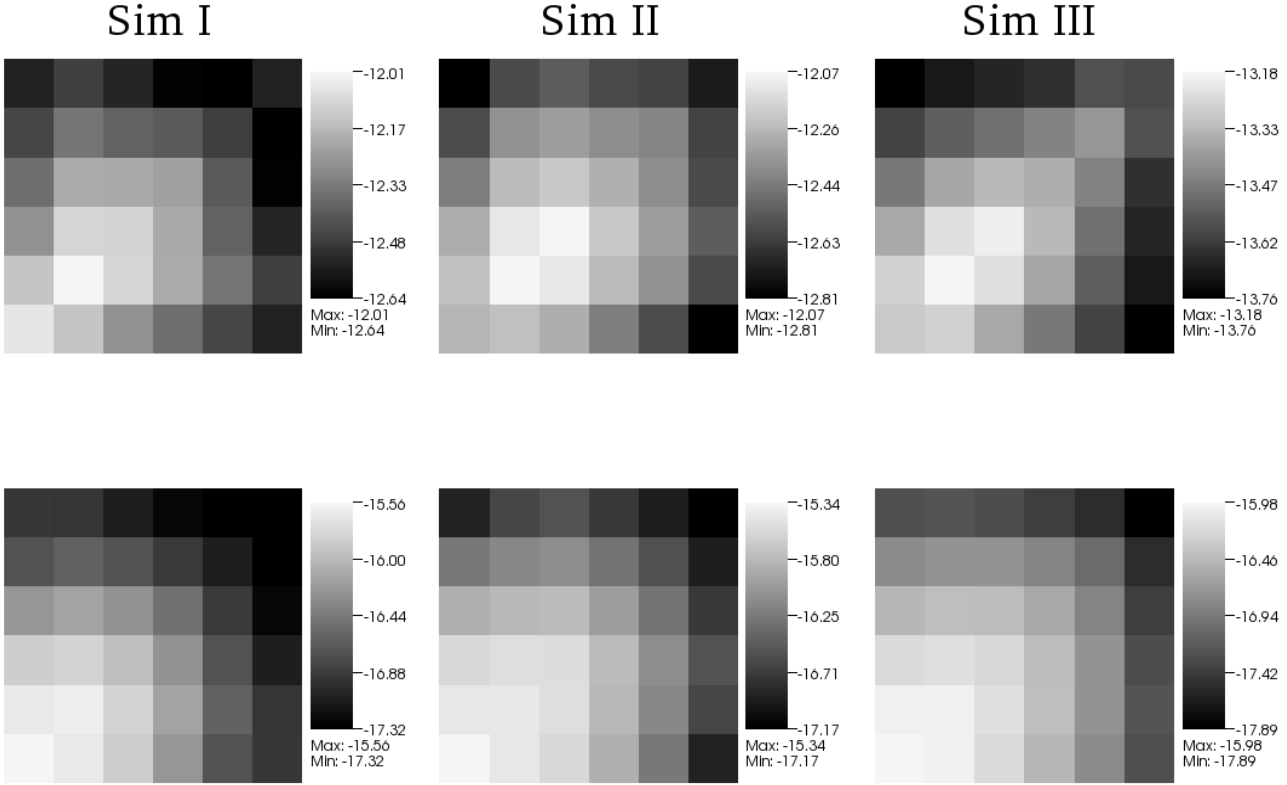,width=.98\textwidth}
\end{tabular}
\caption{\label{fig:matrix} The upper row of plots shows the logarithm of the amplitude of the  $\langle M_{\rm ap}^2\rangle(\theta)$ covariance matrix for the six angular bins we use for our cosmological interpretation. From the left to the right  we show the results for the set of simulations I, II and III.   The bottom row shows the logarithm of the amplitude of the  $\langle M_{\rm ap}^3\rangle(\theta)$ covariance matrix. In all cases we rescaled the amplitude of the covariance matrix to the size of the COSMOS survey as indicated in the text.  }
\end{figure*}
 
\begin{figure*}
\begin{tabular}{ll}
\psfig{figure=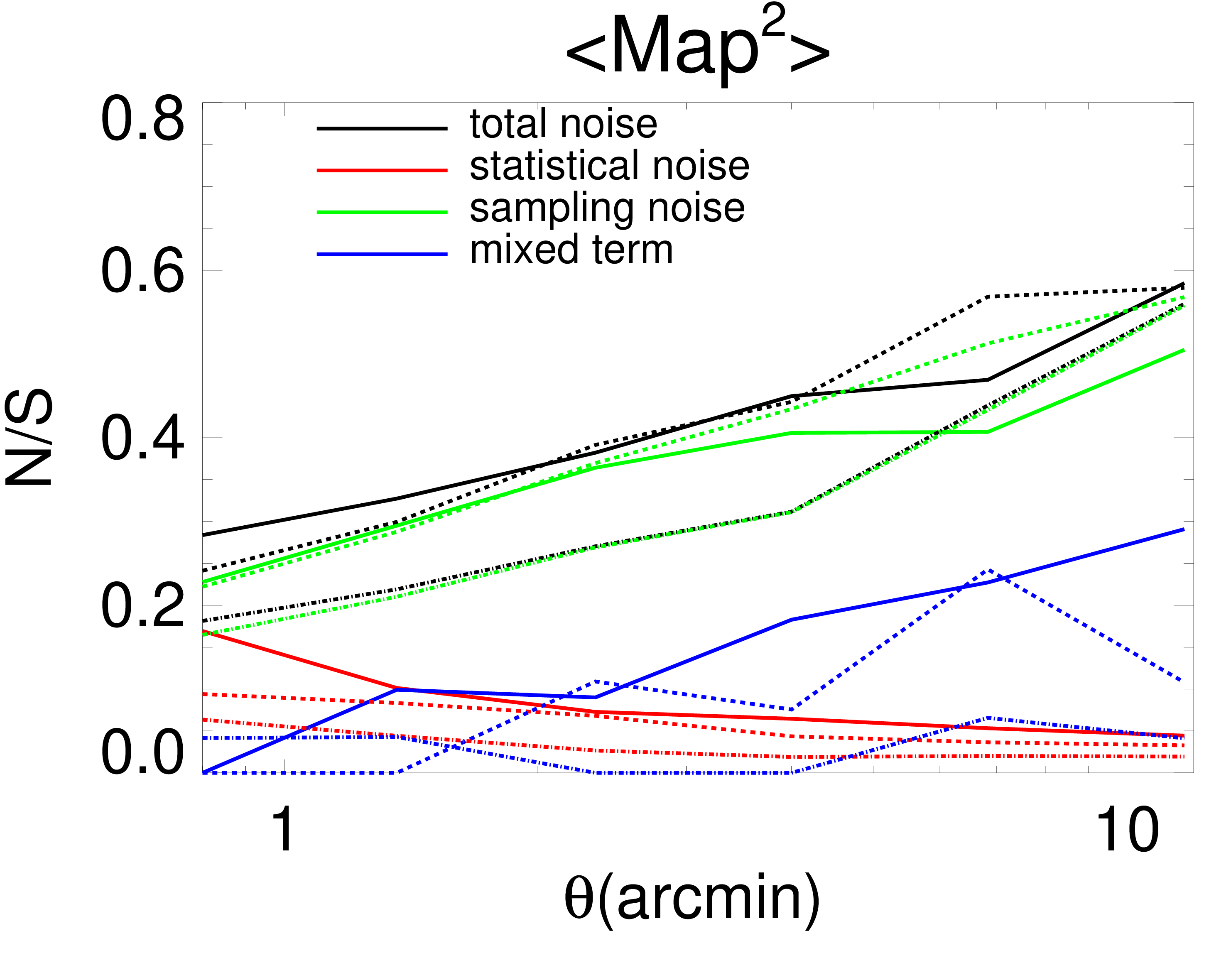,width=.48\textwidth}&\psfig{figure=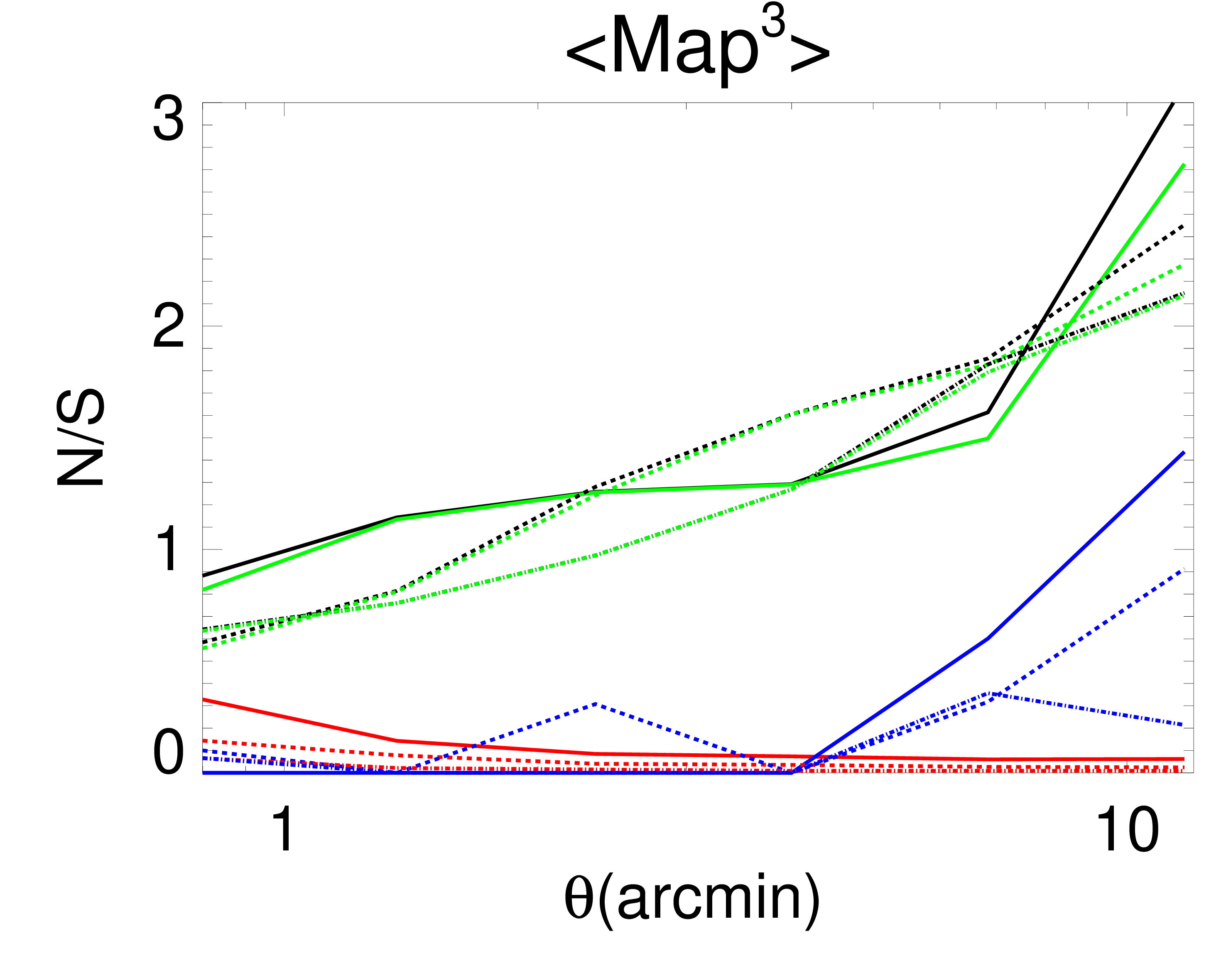,width=.48\textwidth}
\end{tabular}
\caption{\label{fig:covariance} Left panel: noise-to-signal ratio expected for the $\langle M_{\rm ap}^2\rangle(\theta)$ measurement on the COSMOS survey, estimated using the set of simulations I (solid lines) II (dashed lines) and III (dot-dashed lines). The solid black lines show the total noise-to-signal ratio, whereas the green lines show the amplitude of the noise-to-signal coming from sampling variance, the red  lines show the noise-to-signal given by shape noise and the blue lines show the noise-to-signal  from the cross term. Right panel: same as left panel for the measurement of $\langle M_{\rm ap}^3\rangle(\theta)$.} 
\end{figure*}

We notice that our result for $\langle M_{\rm ap}^2 \rangle(\theta)$ is in agreement at the $1 \sigma$ level with the result  \mbox{$\sigma_8 \Big(\frac{\Omega_m}{0.30}\Big)^{0.48}=0.81 \pm 0.17$}  found by Massey et al. (2007b) who previously measured two-point shear statistics on the COSMOS catalogue performing an analysis which is completely independent from the one presented in this paper. 
Our results are also in excellent agreement with the 2D analysis result \mbox{$\sigma_8 \Big(\frac{\Omega_m}{0.30}\Big)^{0.62}=0.68 \pm 0.11$} presented by Schrabback et al. (2010)  who used the same catalogue. The very small difference between the results may be originated from the fact that Schrabback et al. (2010)
used the correlation function to perform a likelihood analysis and a different type of binning which effectively probes the power spectrum at slightly different scales. In addition, they quoted marginalised means, whereas our fit describes the maximum posterior, which typically corresponds to slightly higher $\sigma_8$.

% in our analysis we include all objects, whereas Schrabback et al. (2010) excluded the faint low redshift objects (i.e. with photometric redshift $z<0.6$ and $i^+>24$) due to potential contamination with high redshift galaxies. If among these galaxies there are high redshift outliers, then our redshift distribution has a mean redshift which is smaller than the real one, with a consequent overestimation of $\sigma_8$.  %As a further difference they quoted marginalised means, whereasour fit describes the maximum posterior, which typically corresponds to slightly higher $\sigma_8$. 

 The likelihood analysis of $\langle M_{\rm ap}^3\rangle(\theta)$ gives contours which are fully consistent with the ones obtained using $\langle M_{\rm ap}^2 \rangle(\theta)$, although they are broader. The best-fit cosmology for $\langle M_{\rm ap}^2 \rangle(\theta)$ and $\langle M_{\rm ap}^3\rangle(\theta)$ is  similar suggesting that our measurement is not affected by large systematics which, in principle should affect the two- and three-point statistics in a different way.

To investigate the effect of poor modelling  on the cosmological constraints we just derived, we repeat the likelihood analysis using as data vector the measurement of  $\langle M_{\rm ap}^2\rangle(\theta)$ and $\langle M_{\rm ap}^3\rangle(\theta)$ measured on the Millennium ray-tracing simulations (i.e. set II).  For this test we change our  WMAP7 priors  to the  ones in the Millennium simulation. We find  that for $\langle M_{\rm ap}^2\rangle(\theta)$ the best-fit is $\sigma_8=0.92^{+0.09}_{-0.12}$  for $\Omega_m=0.25$, whereas for $\langle M_{\rm ap}^3\rangle(\theta)$ the best-fit is $\sigma_8=0.95^{+0.09}_{-0.23}$.   The difference between the best-fit values of $\sigma_8$ obtained from the likelihood analysis of  $\langle M_{\rm ap}^2\rangle(\theta)$ and $\langle M_{\rm ap}^3\rangle(\theta)$ is marginal and the inaccuracy of the cosmological predictions  should not bias significantly our results.

\subsection{Effect of the cosmology on the likelihood analysis}\label{sec:cosmodep}

 The result in the previous section have been obtained using the covariance matrix from the Millennium Simulation. It has been pointed out by Eifler, Schneider \& Hartlap (2009) that in the case of two-point shear statistics, the cosmology used to compute the covariance affects the error on the cosmological constraints. In fact, we expect the effect to be even stronger for the three-point shear statistics given their sensibility to $\sigma_8$.
Eifler, Schneider \& Hartlap (2009) suggest to compute the likelihood in each point of the parameter space by using the covariance matrix obtained from the cosmology characterised by those same parameters. This would require us to have large sets of ray-tracing simulations for each point in the parameter space we want to explore, which we do not have.
 However, since we have three sets of simulations which explore a quite large range of $\sigma_8$ values, we can use all of them to compute covariance matrices and evaluate the differences in the likelihood analysis.  For this test, we compute the covariance matrices using the direct measurement of the aperture mass statistics. The covariance matrices obtained in this way reflect less accurately the noise specific to our survey and our estimator,  but they can be computed much faster.
  
   For each set, we create lines-of-sight weighting the convergence maps $\kappa(z_i)$ according to $p_{\rm s}(z)$ given by Equation (\ref{eq:pofz}) to have the same source redshift distribution  as the best-fit to the COSMOS data. 
We account for intrinsic ellipticity dispersion by adding to each pixel of the corresponding convergence noise defined as \cite{vW00,Vaetal09}:
\be\label{eq:noise}
\sigma_\kappa^2= \frac{\sigma_e^2}{2 n \theta^2}
\ee
where $\sigma_e=0.44$, $\theta$ is the pixel size and $n=75/{\rm arcmin}^2$ is the density of galaxies.

From Equation (\ref{eq:filter}) one  notices that the filter $U_\theta(\vartheta)$ has infinite support thus the measurement of $M_{\rm ap}(\theta)$ on a survey of finite size is biased. In practice, for a given characteristic size $\theta$ we can truncate the measurement of $M_{\rm ap}(\theta)$  at a scale $4\theta$.  Because of that, given a simulation of side length $l$ the effective area used to compute the aperture mass  is  $(l-8\theta)^2$ as one cannot compute the aperture mass in regions which are closer than $4\theta$ to the borders. Hence, the effective area used for each bin $\theta$ is different, leading to a covariance matrix which has noise properties which are different for each bin and it depends in a complicated fashion on the angular scale and on the area of the simulations. In order to avoid this issue we do not compute the aperture mass in regions which are closer than  $4\theta_{max}$ to the borders,  where $\theta_{max}$ is the maximum angular bin we measure $\langle M_{\rm ap}^2\rangle(\theta)$ and $\langle M_{\rm ap}^3\rangle(\theta)$.  With this cut  the effective area used for the simulation  is $4.12\, {\rm deg^2} $  for set I,  $5.99\, {\rm deg^2} $ for  the  set II,  $29.68\, {\rm deg^2}$ for set III, whereas the initial area of each set was: $12.84\, {\rm deg^2}$, $16\, {\rm deg^2}$, $49\, {\rm deg^2} $. As one can see the cut we applied reduces the area of each set of simulation  considerably. This increases the noise affecting the covariance matrix but allows us to have an homogeneous measurement of the covariance at all scales  using (almost) the same area of the sky.
  
Once we have the final noisy convergence maps, obtained simply by adding the shape-noise maps to the convergence maps we  estimate for each pixel the value of the aperture mass  using the relation between the convergence $\kappa$  and the aperture mass $M_{\rm ap}$ in Equation (\ref{eq:map_def}). Finally we  compute for each field the second- and third-order moments. We can rescale the final  covariances multiplying by the ratio of the effective area used for  each set of simulations and the area of the COSMOS survey (Schneider et al. 2002). 

In Figure \ref{fig:matrix} we show the amplitude of the elements of the final covariance matrix both for  $\langle M_{\rm ap}^2\rangle(\theta)$ and $\langle M_{\rm ap}^3\rangle(\theta)$ in the six angular bins we  use for the cosmological interpretation. As one can see the overall shape is the same for the three  simulations. However, as one would expect, the overall amplitude of the covariance matrix is larger when the value of $\sigma_8$ is higher. In Figure \ref{fig:covariance} we show the noise-to-signal ratio for each of the terms in the covariance matrix we just computed, i.e. the amplitude of the diagonal part of $\sqrt{{\mathcal C_{ss}}}$, $\sqrt{{\mathcal C_{sn}}}$ and $\sqrt{{\mathcal C_{nn}}}$ over the cosmic shear signal for the set of simulations I, II and III. The plots indicate that for a narrow and deep survey as COSMOS the error generated by sampling variance ${\mathcal C_{ss}}$ is always dominant, as the contribution of statistical noise is minimised by the high density of galaxies. 
In particular, the amplitude of the diagonal part of $\sqrt{{\mathcal C_{ss}}}$ for $\langle M_{\rm ap}^3\rangle(\theta)$ is already as large as the signal at small scales.

%It is interesting to see that, the signal-to-noise ratio does not seem to depend strongly on the cosmology. In fact,  Figure \label{fig:covariance} shows that the noise-to-signal ratio does not depend strongly  both for  $\langle M_{\rm ap}^2\rangle(\theta)$ and  $\langle M_{\rm ap}^3\rangle(\theta)$. 

\begin{figure*}
\begin{tabular}{ll}
\psfig{figure=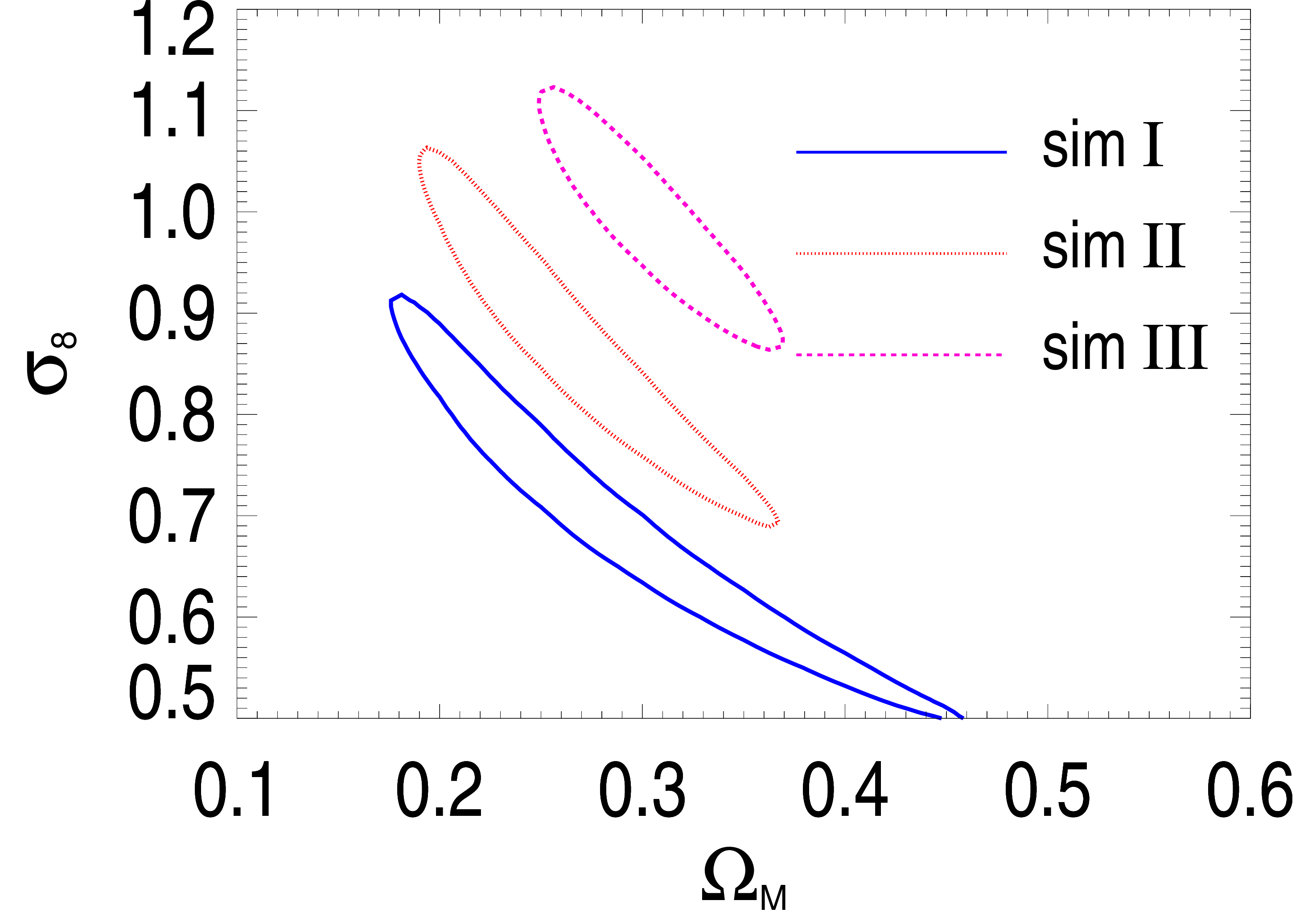,width=0.45\textwidth}&\psfig{figure=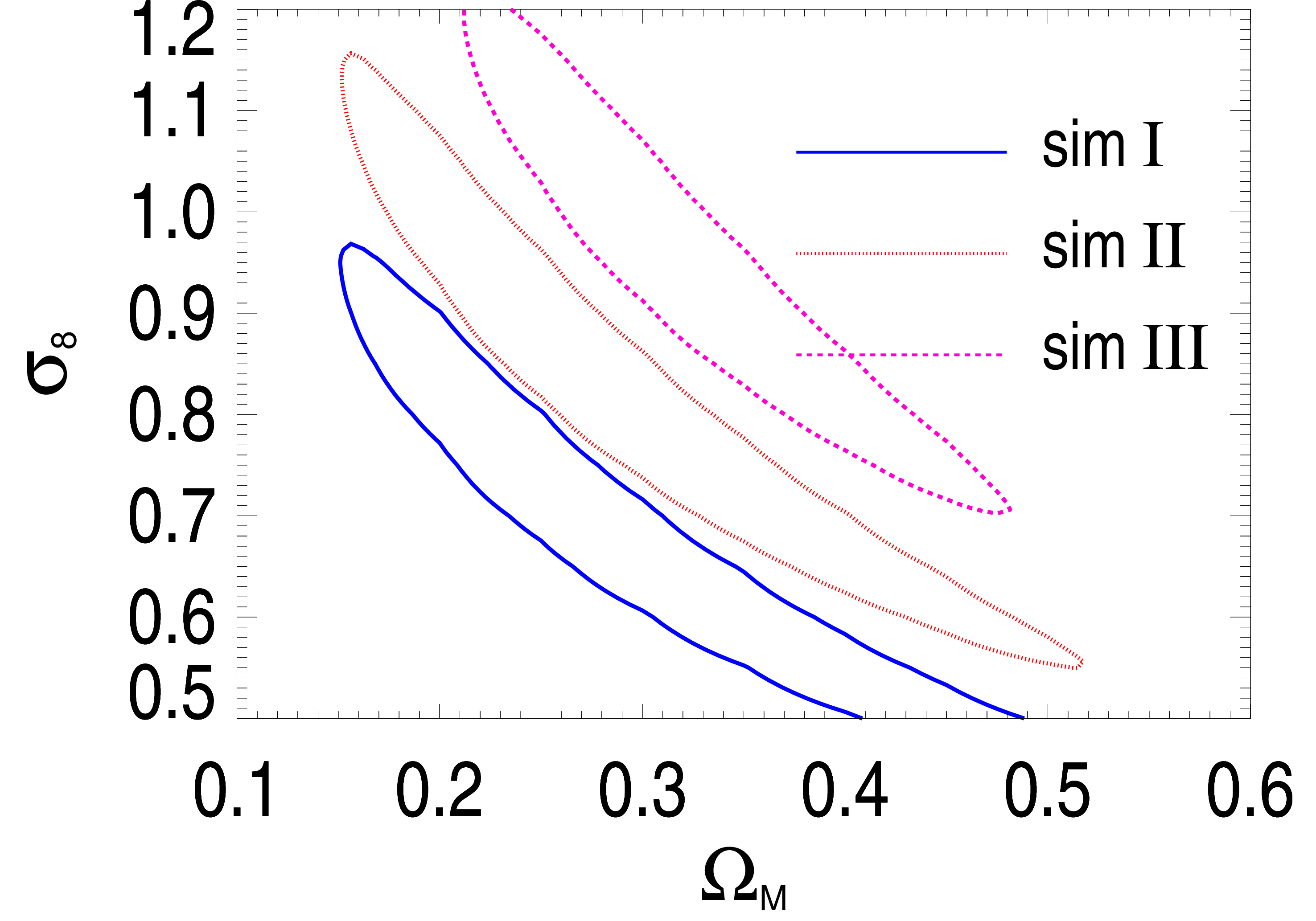,width=0.45\textwidth}
\end{tabular}
\caption{\label{fig:cosmo_map3} The left plot shows  the one sigma constraints in the $[\Omega_m$,$\sigma_8]$ space obtained  for a 20 deg$^2$ simulated survey with the same depth as COSMOS using $\langle M_{\rm ap}^2 \rangle(\theta)$ and  the covariance matrix from the set of simulations I (blue solid contours),  II (red  dotted contours) and III (pink dashed contours). To perform the three likelihood analyses we used the fiducial model which corresponds to the cosmological model employed to  produce each of the set of simulations. The right plot shows the same results for $\langle M_{\rm ap}^3 \rangle(\theta)$.}
%\vspace{2cm}
%\end{minipage}
\end{figure*}

Figure \ref{fig:cosmo_map3} shows the size of the likelihood contours obtained for each set of simulations for  $\langle M_{\rm ap}^2\rangle(\theta)$ (left panel) and  $\langle M_{\rm ap}^3\rangle(\theta)$ (right panel). For each likelihood analysis we used as a data-vector a fiducial model with the same cosmology used to create the simulations. One can see that especially in the case of $\langle M_{\rm ap}^3\rangle(\theta)$ the contours are similar.  The signal-to-noise does not depend strongly on the cosmology. This implies that the covariance matrix strongly depends on the  cosmology. Indeed, the sampling variance term depends very roughly on $\sigma_8^8 \Omega_m^6$ as it is proportional to the signal squared  (compare Equation \ref{eq:map3}). Furthermore, a higher $\sigma_8$ value indicates that, for the same redshift distribution, the structures are more evolved thus the modes are more coupled and the measurements are more correlated.  Since the choice of the fiducial cosmology  has a large impact on the cosmological constraints, therefore one should choose a fiducial cosmology as realistic as possible.  We quantify the impact of the choice of the covariance matrix  in Table \label{tab:1}   showing  the error-bars one would obtain performing a likelihood analysis of the COSMOS data using the covariances I, II, III.  A posteriori, one can see that it is reasonable to use the set I or II to interpret our data as the observations rule out the cosmology of the set III.  
Furthermore, our analysis  confirms that  it is important for future missions, such as Euclid and JDEM, to  be able to perform a likelihood analysis, varying the  covariance matrix according to the parameters as already suggested by  Eifler, Schneider \& Hartlap (2009).

\begin{table}

 \begin{minipage}[b]{0.5\linewidth}
\centering
 \vspace{0.2cm}
\begin{tabular}{ccc}
& $\langle M_{\rm ap}^2 \rangle$ & $\langle M_{\rm ap}^3 \rangle$ \\
\hline
Sim I & $\sigma_8(\Omega_m=0.3)=0.68\pm 0.10$ & $\sigma_8=0.75^ {+0.10}_{-0.16}$ \\
Sim II & $\sigma_8(\Omega_m=0.3)=0.71 ^{+0.10}_{-0.11}$ & $\sigma_8=0.79^{+0.11}_{-0.21}$ \\
Sim III & $\sigma_8(\Omega_m=0.3)=0.69 \pm 0.12$ & $\sigma_8=0.75^{+0.12}_{-0.35}$ 
 \end{tabular}
\vspace{0.8cm}
\end{minipage}
\hspace{0.25cm}
\begin{minipage}[b]{0.45\linewidth}
\centering
\vspace{.2cm}
\caption{\label{tab:1} The table on the left shows the best-fit values and $\pm1\sigma$ error-bars for the $\sigma_8$ parameter for a fixed $\Omega_m=0.3$, obtained performing the same likelihood analysis for the COSMOS data but with covariance matrices derived using different cosmologies. For this comparison we used covariance matrices which have been computed directly from  the moments of the aperture mass statistics (see Section \ref{sec:cosmodep}).  Notice that because of that the  $\pm1\sigma$ error-bars on  $\sigma_8$ are slightly  different from the ones obtained using the covariance matrix computed from the correlation functions.}
\end{minipage}
\end{table} 

%%\begin{table}\label{tab:1}
%\end{table}

%\end{figure}

\subsection{Measurement of $\langle M_{\rm ap}^3 \rangle(\theta_1,\theta_2,\theta_3)$}\label{sec:multi}
\begin{figure*}
\begin{tabular}{lll}
\psfig{figure=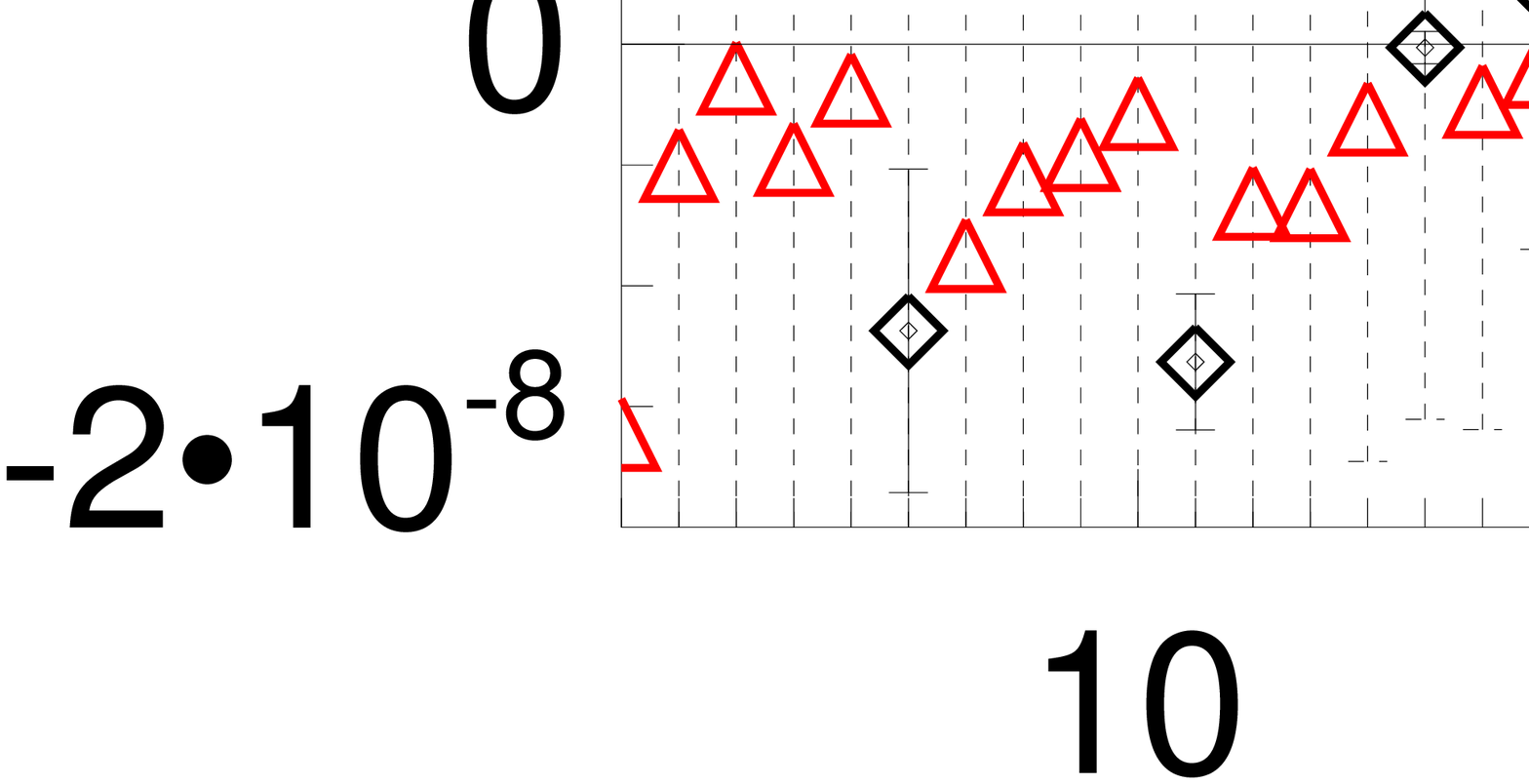,width=.3\textwidth}&\psfig{figure=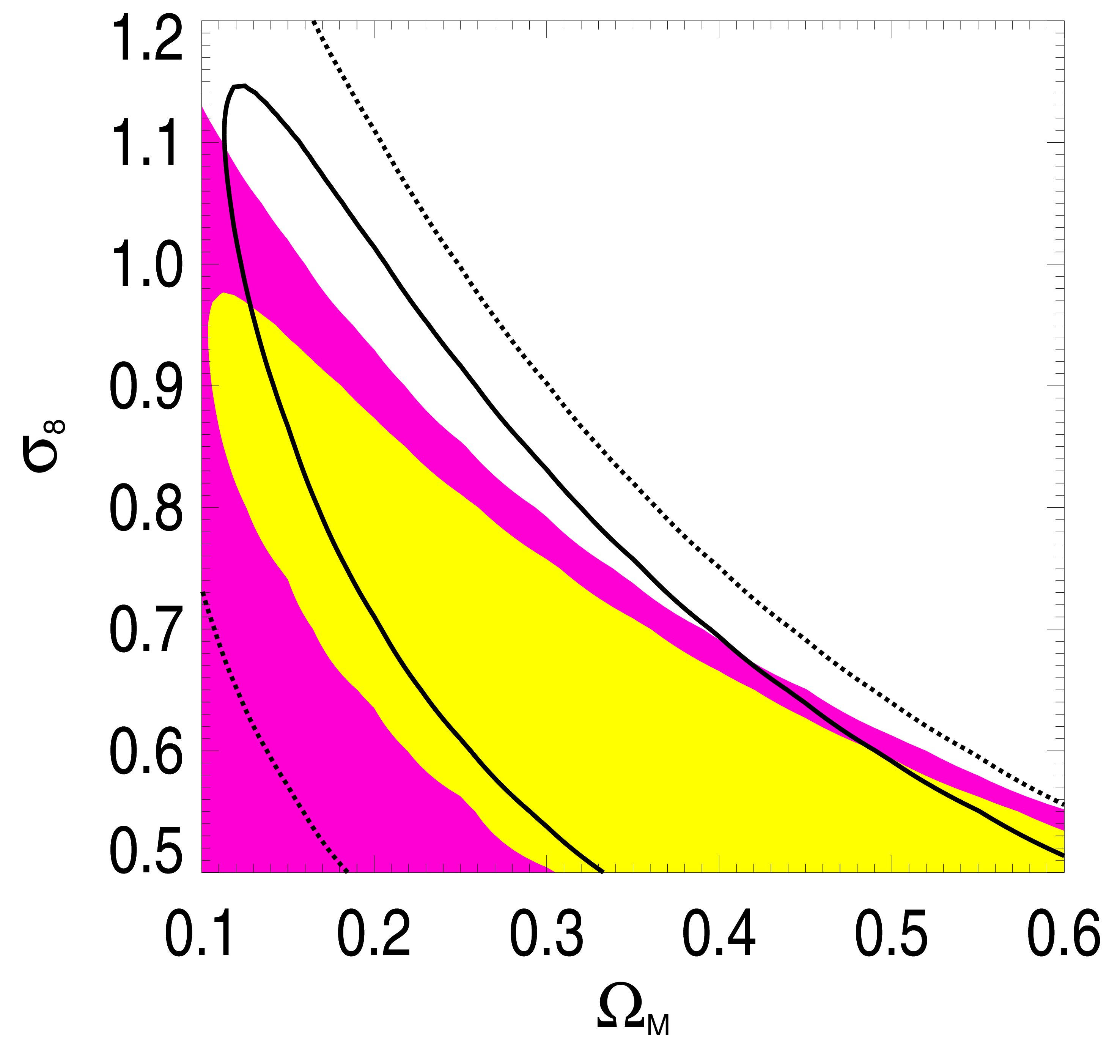,width=.3\textwidth}&\psfig{figure=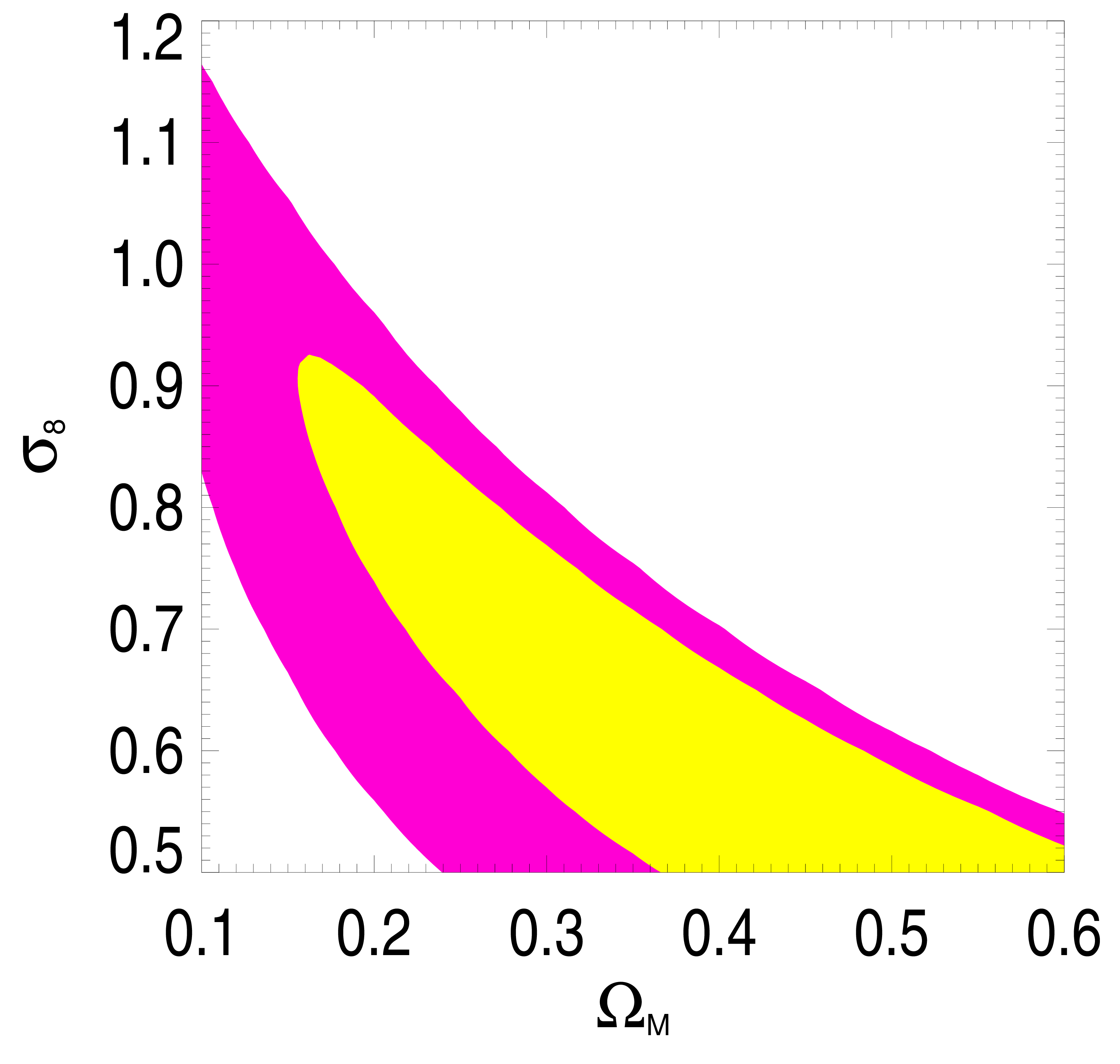,width=.3\textwidth}
\end{tabular}
\caption{\label{fig:map_multi} Left panel: the black diamonds show the amplitude of the cosmological signal $\langle M_{\rm ap}^3 \rangle(\theta_1,\theta_2,\theta_3)$ measured for each of the 56 combinations of $\theta_1$, $\theta_2$ and $\theta_3$ as a function of an identification number which we assigned to each triplet (see table \ref{tab:1}). The amplitude of the cosmological signal is compared with the fiducial WMAP7 cosmology (pink solid line) and the non-gravitational component $\langle M_{\rm ap} M_{\times}^2\rangle(\theta_1,\theta_2,\theta_3)$ is also shown (red diamonds). The solid error-bars for $\langle M_{\rm ap}^3 \rangle(\theta_1,\theta_2,\theta_3)$ include only shape-noise and we omitted the identical error-bars for the non-gravitational component for clarity. The dashed error-bars represent the total errors computed using the ray-tracing set II. Middle panel: likelihood analysis contours obtained using the measurement of $\langle M_{\rm ap}^3 \rangle(\theta_1,\theta_2,\theta_3)$. The black solid (dashed) contours show the likelihood $68\%$ ($95\%$) contours obtained using  $\langle M_{\rm ap}^2 \rangle(\theta)$ for comparison. Right panel: likelihood analysis contours obtained using the measurement of $\langle M_{\rm ap}^3 \rangle(\theta_1,\theta_2,\theta_3)$ combined with $\langle M_{\rm ap}^2 \rangle(\theta)$.}
\end{figure*}

In the previous section we presented the cosmological constraints obtained measuring $\langle M_{\rm ap}^3\rangle(\theta)$ on the COSMOS field and we commented that these constraints are weak due to the small size of the data sample. However, we ignored that the measurement of $\langle M_{\rm ap}^3\rangle(\theta)$ contains only one part of the information about the shape of the bispectrum. Indeed, as it has been pointed out by Schneider et al. (2005), the amplitude of $\langle M_{\rm ap}^3\rangle(\theta)$ mostly depends on the shape of the bispectrum $B(|{\bf k_1}|,|{\bf k_2}|,|{\bf k_2}- {\bf k_1}|)$ for modes with $|{\bf k_1}| \sim |{\bf k_2}| \sim |{\bf k_2}- {\bf k_1}| \sim 1/ \theta $, i.e. on equilateral triangles. One can easily generalise Equation (\ref{eq:map3}) using filters with different characteristic size defining $\langle M_{\rm ap}^3 \rangle(\theta_1,\theta_2,\theta_3)$. By varying independently the three characteristic filter scales one explores the amplitude of the bispectrum measured on a larger variety of triangles. In this section we aim to improve the accuracy of the cosmological constraints obtained in the previous section measuring $\langle M_{\rm ap}^3 \rangle(\theta_1,\theta_2,\theta_3)$ varying $\theta_1$, $\theta_2$ and $\theta_3$ independently. Each angle can take six values 
 between 0.8 arcmin and 12 arcmin with the condition $\theta_1 \leq \theta_2 \leq \theta_3$ as the measurement is symmetric toward any permutation of the filter scales. In this way we obtain 56 measurements of $\langle M_{\rm ap}^3 \rangle(\theta_1,\theta_2,\theta_3)$. Table 2 presents the way in which we order the various triplets of filter scales and the left panel of Figure \ref{fig:map_multi} shows the signal we measured for the triplets. In order to check the cosmological origin of the signal, we divided the signal into E-, B- and mixed-modes.

The amplitude of the E-modes signal is in very good agreement with the predictions. Moreover, the amplitude of the non-gravitational component $\langle M_{\rm ap} M_{\times}^2 \rangle(\theta_1,\theta_2,\theta_3)$ is generally low and consistent with zero in most of the bins, confirming that this estimator is not substantially affected by residual PSF systematics and can be used to constrain cosmology.

Ideally we would like to include the whole set of new measurements in the likelihood analysis. However, we have a small set of simulations, so the covariance matrix may be poorly conditioned and lead to biased constraints if we choose too many measurements or measurements which are highly correlated. 

%In order to identify which set of triplets $(\theta_1,\theta_2,\theta_3)$ are highly correlated, we define the correlation matrix:
%\begin{equation} 
%{\mathcal Corr} (x_i,x_j)= \frac{{\mathcal C}(x_i,x_j)}{\sqrt{{\mathcal C}(x_i,x_i)}\sqrt{{\mathcal C}(x_j,x_j)}}
%\end{equation}
%where $x_i$, $x_j$ are the measurements of $\langle M_{\rm ap}^3 \rangle(\theta_1,\theta_2,\theta_3)$ for two different triplets $i$ and $j$.
%We notice that the points are very correlated. This is not unexpected, as the three-point shear statistics are generated by non-linear evolution of the matter density field, and in this regime modes are strongly coupled. Using a fiducial model, we find that a likelihood analysis of this quantity using the covariance matrices we have is highly unstable.

The $56\times56$ covariance matrix  has  a condition number, i.e. the ratio between the highest and the smallest eigenvalues, of $10^{15}$ indicating that as long as we keep all the scales the inversion is unstable. In fact, with such a high condition number every modelling error will have a huge impact on the likelihood result. Note that since we use a logarithmic step, the 56 triplets are mainly composed by triplets where  $\theta_1$, $\theta_2$ and $\theta_3$  are similar and small. These triplets  explore the bispectrum at very similar modes. Hence, their measurement is highly correlated.

In order to reduce  the number of triplets  having similar filter scales, we retain only  triplets with  $\theta_1=\theta_2$ if  $\theta_1<4\, {\rm arcmin}$. If $\theta_1>4\,{\rm arcmin}$  the filter sizes are very different so the triplets can be all retained.  With these conditions we are left with 25 triplets. The condition number for the new $25\times 25$ covariance matrix  is  $10^3$, indicating that the likelihood analysis with these triplets is well-conditioned.
Hence, we perform a likelihood analysis using these 25 $\langle M_{\rm ap}^3 \rangle(\theta_1,\theta_2,\theta_3)$ measurements from the COSMOS data. We show  in the middle panel of Figure \ref{fig:map_multi} the posterior probability distribution for $\Omega_m$ and $\sigma_8$ after marginalisation over the redshift parameter $z_0$.  One can see that the cosmological constraints are significantly improved. The  best-fit \mbox{$\sigma_8=0.69^{+0.08}_{-0.14} $} for $\Omega_m=0.3$ is in good agreement with the constraints we found using  $\langle M_{\rm ap}^2 \rangle (\theta)$ and $\langle M_{\rm ap}^3 \rangle (\theta)$.
\begin{table}
\begin{minipage}[b]{0.5\linewidth}
\centering
\begin{tabular}{cccc}
$\theta_1$ &$\theta_2$ & $\theta_3$ & number \\
\hline
\hline
$\theta_1$ & $\theta_1$ & $\theta_1$ & $1$\\
$\theta_1$ & $\theta_1$ & $\theta_2$ & $2$\\
$\cdots$ & $\cdots$ & $\cdots$ & $\cdots$ \\
$\theta_1$ & $\theta_1$ & $\theta_6$ & $6$\\ 
$\theta_1$ & $\theta_2$ & $\theta_2$ & $7$\\ 
$\cdots$ & $\cdots$ & $\cdots$ & $\cdots$ \\
$\theta_2$ & $\theta_2$ & $\theta_2$ & $12$\\ 
$\cdots$ & $\cdots$ & $\cdots$ & $\cdots$ \\
$\theta_6$ & $\theta_6$ & $\theta_6$ & $56$
\end{tabular}
\end{minipage}
\begin{minipage}[b]{0.4\linewidth}
\centering
\vspace{2cm}
\caption{\label{tab:1} The first three columns show the values taken by the first, second and third aperture mass angles: $\theta_1$, $\theta_2$, $\theta_3$, where $[\theta_1, \cdots,\theta_6]$ indicate six angular logarithmic bins between 0.8 arcmin and 12 arcmin. The last column shows the number associated to each triplet shown in  the left panel of Figure \ref{fig:map_multi}.}
\end{minipage}
\end{table}
In the right panel of Figure \ref{fig:map_multi}, we show that the combined analysis of  $\langle M_{\rm ap}^2(\theta)$  and  $\langle M_{\rm ap}^3 \rangle(\theta_1,\theta_2,\theta_3)$ increases the strength of the cosmological constraints.  Indeed, the combined likelihood results yields  \mbox{$\sigma_8 (\Omega_m/0.30)^{0.50}=0.69 ^ {+0.07}_{-0.12}$}. %This because the degeneracy direction of $\langle M_{\rm ap}^2(\theta)$  and  $\langle M_{\rm ap}^3 \rangle(\theta_1,\theta_2,\theta_3)$  is different. 

Using  Fisher matrix analysis, various authors \cite{KiSc05,TaJa05,Beetal10} showed that $\langle M_{\rm ap}^3 \rangle(\theta_1,\theta_2,\theta_3)$ contains more information than $\langle M_{\rm ap}^3 \rangle(\theta)$.  They also suggested that since $\langle M_{\rm ap}^2\rangle(\theta)$  and  $\langle M_{\rm ap}^3 \rangle(\theta_1,\theta_2,\theta_3)$  depend differently on $\Omega_m$ and $\sigma_8$ they can be combined to partially break the $\sigma_8-\Omega_m$ degeneracy.  In this respect our results are in good agreement with their prediction and show the potential associated to this measurement.

 %However the cosmological constraints we show in this paper using  $\langle M_{\rm ap}^3 \rangle(\theta_1,\theta_2,\theta_3)$ cannot be considered as accurate as the error-bars generated by the likelihood analysis suggests. We already discussed how the cosmological constraints we obtained are sensitive to the non-linear evolution of the power spectrum.  Various aspects which have been neglected in this paper are likely to affect our results. For example, we did not account for massive neutrinos  and the effect of baryons on small angular scales: both are expected changes the evolution of the matter fluctuations in particular in non-linear regime. Furthermore, we did not account for lens-clustering effects which also are expected to change the  amplitude of the measured three-point shear statistics \cite{Haetal02}.

\section{Conclusions}\label{sec:conclusions}

 In the past decade several papers  discussed the use of  the three-point statistics  to explore the non-linear features of the field of matter fluctuations. It has also been pointed out that adding three-point to the two-point shear measurements allows one to increase the accuracy of the constraints on cosmological parameters (see Vafaei et al. 2010 for a quantitative analysis based on recent ray-tracing simulations). However, measuring three-point shear statistics appears much more challenging and only few attempts have been done so far. The aim of this paper was to show that today we have the expertise and the data quality to allow for the measurement of third-order cosmic shear statistics. For such a purpose, we used the COSMOS data set: a mosaic of contiguous pointings of space-based data which represents the ideal benchmark to test the quality of weak lensing analyses for future space-based data. In order to show the reliability of our results we focused our attention on some very important aspects to correctly interpret three-point shear statistics.

As first step, we tested the accuracy of the cosmological predictions against four sets of recent ray-tracing simulations based on ${\rm \Lambda CDM}$ cosmologies. We found that current cosmological predictions for the two- and three-point shear statistics used in this paper, i.e. $\langle M_{\rm ap}^2 \rangle(\theta)$ and $\langle M_{\rm ap}^3 \rangle(\theta)$, are able to reproduce the signal measured in the simulations typically only within $10\%$ and $20\%$ accuracy respectively. This clearly demonstrates that improved model predictions will be required for the analysis of future surveys.

We used empirical models derived using $N$-body simulations to quantify the effect of the shape-shear coupling on the measurement of the two- and three-point shear statistics. We found that for the COSMOS Survey this effect should be subdominant. To date, the shear-shape effect is still poorly constrained and the empirical models we used are based on a toy-model which might not be realistic enough. More work still needs to be done to understand 
and model the effect of the intrinsic orientation of galaxies on cosmic shear statistics,  although the recent  results by  Mandelbaum et al. (2009) suggest that the shear-shape  should affect the value of  $\sigma_8$ from two-point shear statistics at most at the $2\%$ level.

In order to demonstrate the robustness of our measurement, we divided the signal into gravitational and non-gravitational components and investigated the amplitude of the cross-correlations between galaxies and stars, which can be used to reveal the existence of PSF residual correlations. We found that the measured $\langle M_{\rm ap}^3\rangle(\theta)$ signal has an amplitude which is in very good agreement with the WMAP7 best-fit model, and that all conducted tests are consistent with negligible residual PSF systematics. Moreover, we showed that the non-gravitational component is consistent with zero both for the two- and the three-point shear statistics. The cross-correlation between stars and galaxies is also consistent with zero. 

The likelihood analysis of $\langle M_{\rm ap}^3\rangle (\theta)$ gives $\sigma_8=0.78^{+0.11}_{-0.26}$ for fixed \mbox{$\Omega_m=0.30$}.
The likelihood results obtained using $\langle M_{\rm ap}^3\rangle(\theta) $ alone are in very good agreement with the ones obtained using $\langle M_{\rm ap}^2\rangle(\theta) $. Our error budget is dominated by the large sampling variance which strongly  affects the measurement of the three-point shear statistics on a small field  such as COSMOS. In fact, joining the two measurements we obtain the same cosmological constraints derived by using $\langle M_{\rm ap}^2\rangle(\theta)$ alone. This is in agreement with the results by Vafaei et al. (2010), who found that a combined measurement of  $\langle M_{\rm ap}^2\rangle(\theta)$ and $\langle M_{\rm ap}^3\rangle(\theta)$  on narrow deep surveys such COSMOS cannot be used effectively to break the $[\Omega_m,\sigma_8]$ degeneracy.  Finally, we generalised the definition of $\langle M_{\rm ap}^3\rangle(\theta)$ to $\langle M_{\rm ap}^3 \rangle(\theta_1,\theta_2,\theta_3)$ and we showed that its amplitude is in excellent agreement with the WMAP7 best-fit model.  We find, as expected, that this measurement improves the accuracy of the cosmological constraints  $\sigma_8=0.69^{+0.08}_{-0.14}$ for fixed \mbox{$\Omega_m=0.30$}.  The combined analysis of  $\langle M_{\rm ap}^2(\theta)$  and  $\langle M_{\rm ap}^3 \rangle(\theta_1,\theta_2,\theta_3)$ further increases the strength of the cosmological constraints.  Indeed, the combined likelihood results yields  \mbox{$\sigma_8 (\Omega_m/0.30)^{0.50}=0.69 ^ {+0.07}_{-0.12}$}. 

We find this result very encouraging, however we would like to make clear once more that  the lack of precision which affects the non-linear evolution modelling, is today still  significant. We already discussed how the cosmological constraints we obtained are sensitive to the non-linear evolution of the power spectrum. Note that Lawrence et al. (2010) very recently provided a substantially improved prescription for the non-linear power spectrum , but similar results will also be required for the bispectrum.   Various aspects which have been neglected in this paper are likely to affect our results. For example, we did not account for massive neutrinos  and the effect of baryons on small angular scales: both are expected to change the evolution of the matter fluctuations in particular in the non-linear regime. Furthermore, we did not account for lens-clustering effects which are also expected to change the  amplitude of the measured three-point shear statistics \cite{Haetal02}.

Constraints from wide surveys will more strongly benefit from the inclusion of third-order statistics.  Hence, for future surveys such as Euclid, LSST and JDEM covering thousands of square degrees it will be possible to use the measurement of the two- and three-point shear statistics together with photometric redshift information to infer tight cosmological constraints. In this paper we did not make use of the photometric redshifts estimated  for individual sources; we expect the tomographic measurement of the third-order shear statistics to add further information and improve the precision of the cosmological constraints.

In the prospective of these  future weak-lensing missions,  it is important to show that it is today possible to measure three-point shear statistics, to quantify the level of systematics and to interpret the cosmological origin of the signal. It is also important to point out which are the limiting factors  requiring further work. This paper shows that the measurement and interpretation of third-order statistics are possible and have high potential; our detection is in very good agreement with the WMAP7 best fit-model and this together with the fact that our measurement is robust against systematics, is a very encouraging outcome and an important proof of concept for future weak lensing missions.

%The lack of sufficiently accurate cosmological predictions and the poor knowledge of the shear-shape effect on three-point shear statistics are not a limitation for the goal of our paper to provide an important proof of concept for future weak lensing missions such as Euclid, JDEM and LSST. %In addition, our errors are dominated by sampling variance due to the small area of the COSMOS survey.
\section*{Acknowledgments}
This work is based on observations made with the NASA/ESA {\it Hubble Space Telescope}, obtained from the data archives at the Space Telescope European Coordinating Facility and the Space Telescope Science Institute. 
The authors would like to thank Aaryn Tonita, Henk Hoekstra, Peter Schneider and Ismael Tereno for useful discussions and for reading carefully our manuscript.
We also would like to thank the University of British Columbia, where part of this project has been developed. ES acknowledges the support from the  Netherlands Organization for Scientific Research (NWO) through a Vidi  grant (number: 639.042.814)  and  the financial support from the Alexander von Humboldt foundation. TS acknowledges the financial support from the Netherlands Organization for Scientific Research (NWO). TS and JH acknowledge the financial support from 
the Deutsche Forschungsgemeinschaft through SFB/Transregio 33 ``The dark Universe''. LvW acknowledges the support from NSERC, CFI and CIAR. SH acknowledges support by the Deutsche Forschungsgemeinschaft within 
the Priority Programme 1177 under the project SCHN 342/6. It is a pleasure to acknowledge the support of the European Community Research Training Network (DUEL), which facilitated the collaboration between the different institutes.

%\bibliographystyle{mn2e}
%\bibliography{test}

\begin{thebibliography}{}
\bibitem[Bacon et al. 2003]{Baetal03} Bacon D., Massey R., Refregier A., Ellis R., 2003, MNRAS, 344, 673
\bibitem[Benjamin et al. 2008]{Beetal08} Benjamin J. et al., 2008, MNRAS, 381, 702
\bibitem[Berg\'{e}, Amara \& R\'{e}fr\'{e}gier  2010]{Beetal10} Berg\'{e} J., Amara A., R\'{e}fr\'{e}gier, A., 2010, ApJ, 712, 992
\bibitem[Bernardeau, van Waerbeke \& Mellier 1997]{BvWM97} Bernardeau F., van Waerbeke L., Mellier Y., 1997, A\&A, 322, 1
 \bibitem[Bernardeau, van Waerbeke \& Mellier 2003]{Beetal02} Bernardeau F., Mellier Y., van Waerbeke L., 2003, A\&A, 397, 405
\bibitem[Bertin \& Arnouts 1996]{BeAr96} Bertin E., Arnouts S., 1996, A\&A, 117, 393
\bibitem[Brainerd et al. 1996]{Bretal96} Brainerd T., Blandford R.~D., Smail I., 1996, ApJ, 466, 623
\bibitem[Bridle et al. 2009]{Bretal09} Bridle S. et al., 2009, MNRAS submitted, e-prints, arXiv:0908.0945
\bibitem[Crittenden et al. 2001]{Cretal01} Crittenden R., Natarajan P., Pen
 U.~L., Theuns, T., 2001, ApJ, 559, 552
\bibitem[Crittenden et al. 2002]{Cretal02} Crittenden R., Natarajan P., Pen
 U.~L., Theuns, T., 2002, ApJ, 568, 20
%\bibitem[Dunkley et al. 2009] {Duetal09} Dunkley J. et al., 2009, ApJS, 180, 306
\bibitem[Eifler, Schneider \& Hartlap]{Eietal09} Eifler T., Schneider P., Hartlap J., 2009, A\&A, 502, 721
\bibitem[Eisenstein \& Hu 1998]{EiHu98} Eisenstein D.~J., Hu W., 1998, ApJ, 496, 405
\bibitem[Erben et al. 2001]{Eretal01} Erben T., van Waerbeke L., Bertin E., Mellier Y., Schneider P., 2001, A\&A, 366, 717
\bibitem[Fu et al. 2008]{Fuetal08} Fu L. et al., 2008, A\&A, 479, 9
\bibitem[Hamana et al. 2002]{Haetal02} Hamana T., Colombi S., Thion A., Devriendt J., Mellier Y., Bernareadu F., 2002, MNRAS, 330, 365
\bibitem[Hartlap, Simon \& Schneider 2007] {Haetal07} Hartlap J., Simon P., Schneider P., 2007, A\&A, 2007, 464, 399
\bibitem[Hartlap et al. 2009]{Haetal09} Hartlap J., Schrabback T., Simon P., Schneider P., 2009, A\&A, 504, 689
%\bibitem[Heymans 2001]{He01} Heymans C., 2001, PhD thesis.
\bibitem[Heymans et al. 2006a]{Heetal06a} Heymans C. et al., 2006a, MNRAS, 368,1323
\bibitem[Heymans et al. 2006b]{Heetal06b} Heymans C., White M., Heavens A., Vale C., van Waerbeke L., 2006b, MNRAS, 371, 750
\bibitem[Hilbert et al. 2009]{Hietal09} Hilbert S., Hartlap J., White S., Schneider P., 2009, A\&A, 499, 31
\bibitem[Hirata \& Seljak 2004]{HiSe04} Hirata C., Seljak U., 2004, Phys. Rev. D, 70, 6
%\bibitem[Hirata et al. 2007]{Hietal07} Hirata C., Mandelbaum R., Ishak M., Seljak U., Nichol R., Pimbblet K., Ross N., Wake D., 2007, MNRAS, 38, 1197
\bibitem[Hoekstra et al. 1998]{Hoetal98} Hoekstra H., Franx M., Kuijken K., Squires G., 1998, ApJ, 504, 636 
%\bibitem[Hoekstra 2004]{Ho04} Hoekstra, H., 2004, MNRAS, 347, 1337
%\bibitem[Hoekstra et al. 2006]{Hoetal06} Hoekstra H. et al., 2006, ApJ, 347, 116 
%\bibitem[Jain, Jarvis \& Bernstein 2006]{JJB06} Jain B., Jarvis M., Bernstein G., 2006, JCAP, 2, 1
\bibitem[Ilbert et al. 2009]{Iletal09} Ilbert O. et al., 2009, ApJ, 690, 1236
\bibitem[Jarvis, Bernstein \& Jain 2004]{Jaetal04} Jarvis M., Bernstein G. , Jain B., 2004, MNRAS, 352, 338
\bibitem[Joachimi, Schneider \& Eifler 2008]{Joetal08} Joachimi B., Schneider P., Eifler T., 2008, A\&A, 477, 43
\bibitem[Kaiser, Squires \& Broadhurst 1995]{Kaetal95} Kaiser N., Squires J., Broadhurst T., 1995, ApJ, 449, 460
\bibitem[Kaiser 1998]{Ka98} Kaiser N., 1998, ApJ, 498, 26
\bibitem[Kilbinger \& Schneider 2005]{KiSc05} Kilbinger M., Schneider P., 2005, A\&A, 442, 69
\bibitem[Kitching et al. 2009]{Kietal08} Kitching T., Amara A., Abdalla F., Joachimi B., Refregier A., 2009, MNRAS, 399, 2107 
\bibitem[Komatsu et al. 2010]{Koetal10} Komatsu E. et al., 2010, ApJS submitted, e-prints, arXiv:1001.4538
\bibitem[Lawrence et al. 2010]{Laetal09} Lawrence E., Heitmann K., White M., Higdon D., Wagner C., Habib S., Williams B., 2009, ApJ submitted, e-prints, ArXiv:0912.4490
\bibitem[Luppino \& Kaiser 1997]{LuKa97} Luppino G.~A., Kaiser N., 1997, ApJ, 475, 20
\bibitem[Mandelbaum et al. 2006]{Maetal06} Mandelbaum R., Hirata C., Ishak M., Seljak U., Brinkmann J., 2006, MNRAS, 367, 611
\bibitem[Mandelbaum et al. 2009]{Maetal09} Mandelbaum R. et al., 2009, MNRAS submitted, e-prints, arXiv:0911.5347
\bibitem[Massey et al. 2007a]{Maetal07a} Massey R. et al., 2007a, MNRAS, 376, 13
\bibitem[Massey et al. 2007b]{Maetal07b} Massey R. et al., 2007b, ApJS, 172, 239
\bibitem[Peacock \& Dodds 1996]{PD} Peacock J.~A., Dodds S.~J., 1996, MNRAS, 280, 19
\bibitem[Pen et al. 2003]{Peetal03} Pen U.~L., Zhang T., van Waerbeke L., Mellier Y., Zhang P., Dubinski J, 2003, ApJ, 592, 664
\bibitem[Pielorz et al. 2009]{Pietal09} Pielorz J., R\"{o}diger J., Tereno I., Schneider P., 2009, A\&A submitted, e-print, arXiv:0907.1524
\bibitem[Schneider et al. 1998]{Scetal98} Schneider P., van Waerbeke L., Jain B., Kruse G., 1998, MNRAS, 296, 873
\bibitem[Schneider et al. 2002]{Scetal02} Schneider P., van Waerbeke L., Kilbinger M., Mellier Y., 2002, A\&A, 396, 1
\bibitem[Schneider \& Lombardi]{ScLo03} Schneider P., Lombardi M., 2003, A\&A, 397, 809
\bibitem[Schneider, Kilbinger \& Lombardi 2005]{Scetal05} Schneider P., Kilbinger M., Lombardi M., 2005, A\&A, 431, 9
\bibitem[Schrabback et al. 2007]{Scetal07} Schrabback T., Erben T., Simon P., et al., 2007, A\&A, 468, 823
\bibitem[Schrabback et al. 2010]{Scetal10} Schrabback T. et al., 2010, A\&A accepted, e-print ArXiV:0911.0053
\bibitem[Scoville et al. 2007a]{Scoetal07a} Scoville N. et al., 2007,  ApJS, 172, 1
\bibitem[Scoville et al. 2007b]{Scoetal07b} Scoville N. et al., 2007,  ApJS, 172, 38
\bibitem[Scoccimarro \& Couchman 2001]{ScCo} Scoccimarro R., Couchman H.~M.~D., 2001, MNRAS, 325, 1312
\bibitem[Semboloni et al. 2007a]{Seetal07} Semboloni R., van Waerbeke L., Heymans C., Mellier Y., 2007, MNRAS, 375, L6 
\bibitem[Semboloni et al. 2008]{Seetal08} Semboloni E., Heymans C., van Waerbeke L., Schneider P., 2008, MNRAS, 388, 991
\bibitem[Semboloni et al. 2009]{Seetal09} Semboloni E., Tereno I., van Waerbeke L., Heymans C., 2009, MNRAS, 397, 608
\bibitem[Smith et al. 2003]{Smetal03} Smith R.~E. et al., 2003, MNRAS, 341, 1311 
\bibitem[Spergel et al. 2007]{Spetal07} Spergel D. et al, 2007, ApJS, 170, 377
\bibitem[Springel et al. 2005]{Spetal05} Springel V., White S.~D., Jenkins A., et al., 2005, Nature, 435, 629
\bibitem[Takada \& Jain 2005]{TaJa05} Takada M., Jain B., 2005, MNRAS, 348, 897
\bibitem[Takada \& Jain 2009]{TaJa09} Takada M., Jain B., 2009, MNRAS, 395, 2065
\bibitem[Vafaei et al. 2010]{Vaetal09} Vafaei S., Lu T., van Waerbeke L., Semboloni E., Heymans C., Pen U.~L., 2010, Astroparticle Physics, 32, 340
\bibitem[van Waerbeke 2000]{vW00} van Waerbeke L., 2000, MNRAS, 313, 524
\bibitem[van Waerbeke, Bernardeau \& Mellier 1999]{vWetal99} van Waerbeke L., Bernardeau F., Mellier Y., 1999, A\&A, 342, 15
\bibitem[van Waerbeke et al. 2001]{vWetal01} van Waerbeke L., Hamana T., Scoccimarro R., Colombi S., Bernardeau F., 2001, MNRAS, 322, 918
%\bibitem[van Waerbeke et al. 2001b]{vW01} van Waerbeke L. et al., 2001, A\&A, 374, 757
\bibitem[van Waerbeke et al. 2005]{Waetal05} van Waerbeke L., Mellier Y., Hoekstra H., 2005, A\&A, 429, 75
%\bibitem[van Waerbeke et al. 2006]{vWetal06} van Waerbeke L., White M., Hoekstra H., Heymans C., 2006, APh, 26, 91 	
\bibitem[Zhang \& Pen 2005]{ZaPe05} Zhang L.~L., Pen U.~L., 2005, NewAstronomy, 10, 569
\end{thebibliography}

\end{document}